\documentclass[12pt,a4paper]{article}

\usepackage{lineno}  
\usepackage{graphicx}  
\usepackage{xspace} 
\usepackage{color}
\usepackage{colortbl}
\usepackage{amsmath} 
\usepackage{ifthen} 

\usepackage{subfigure}
\usepackage{rotating}
\usepackage{cite}

\newboolean{pdflatex}
\setboolean{pdflatex}{false} 
\newboolean{articletitles}
\setboolean{articletitles}{true} 

\newboolean{uprightparticles}
\setboolean{uprightparticles}{false} 

\usepackage{amssymb}
\usepackage{amsfonts}
\usepackage{upgreek} 
\usepackage{epstopdf}

\usepackage{hyperref}    
\usepackage[all]{hypcap} 

\def\ux85 {UX85\xspace}

\ifthenelse{\boolean{uprightparticles}}%
{

 \def\PDelta      {\ensuremath{\Delta}\xspace}                 
 \def\PXi      {\ensuremath{\Xi}\xspace}                 
 \def\PLambda      {\ensuremath{\Lambda}\xspace}                 
 \def\PSigma      {\ensuremath{\Sigma}\xspace}                 
 \def\POmega      {\ensuremath{\Omega}\xspace}                 
 \def\PUpsilon      {\ensuremath{\Upsilon}\xspace}                 
 
 \def\PB      {\ensuremath{\mathrm{B}}\xspace}                 
                  
 \def\PD      {\ensuremath{\mathrm{D}}\xspace}

 \def\PK      {\ensuremath{\mathrm{K}}\xspace}

 \def\Pi      {\ensuremath{\mathrm{i}}\xspace}

}
{

 \mathchardef\PDelta="7101
 \mathchardef\PXi="7104
 \mathchardef\PLambda="7103
 \mathchardef\PSigma="7106
 \mathchardef\POmega="710A
 \mathchardef\PUpsilon="7107
                  
 \def\PB      {\ensuremath{B}\xspace}                 
                  
 \def\PD      {\ensuremath{D}\xspace}

 \def\PK      {\ensuremath{K}\xspace}

 \def\Pi      {\ensuremath{i}\xspace}

}

\def\kaon  {\ensuremath{\PK}\xspace}
  \def\Kbar  {\kern 0.2em\overline{\kern -0.2em \PK}{}\xspace}

\def\Kz    {\ensuremath{\kaon^0}\xspace}
\def\Kzb   {\ensuremath{\Kbar^0}\xspace}
\def\KzKzb {\ensuremath{\Kz \kern -0.16em \Kzb}\xspace}
\def\Kp    {\ensuremath{\kaon^+}\xspace}
\def\Km    {\ensuremath{\kaon^-}\xspace}

\def\KpKm  {\ensuremath{\Kp \kern -0.16em \Km}\xspace}

  \def\Dbar    {\kern 0.2em\overline{\kern -0.2em \PD}{}\xspace}
\def\D       {\ensuremath{\PD}\xspace}

\def\Dz      {\ensuremath{\D^0}\xspace}
\def\Dzb     {\ensuremath{\Dbar^0}\xspace}
\def\DzDzb   {\ensuremath{\Dz {\kern -0.16em \Dzb}}\xspace}
\def\Dp      {\ensuremath{\D^+}\xspace}
\def\Dm      {\ensuremath{\D^-}\xspace}

\def\DpDm    {\ensuremath{\Dp {\kern -0.16em \Dm}}\xspace}

  \def\Bbar    {\kern 0.18em\overline{\kern -0.18em \PB}{}\xspace}

  \def\Y#1S{\ensuremath{\PUpsilon{(#1S)}}\xspace}

\def\to                 {\ensuremath{\rightarrow}\xspace}

\def\AT#1     {\ensuremath{A_{\mathrm{T}}^{#1}}\xspace}           

\def\C#1      {\ensuremath{\mathcal{C}_{#1}}\xspace}                       
\def\Cp#1     {\ensuremath{\mathcal{C}_{#1}^{'}}\xspace}                    
\def\Ceff#1   {\ensuremath{\mathcal{C}_{#1}^{\mathrm{(eff)}}}\xspace}        
\def\Cpeff#1  {\ensuremath{\mathcal{C}_{#1}^{'\mathrm{(eff)}}}\xspace}       
\def\Ope#1    {\ensuremath{\mathcal{O}_{#1}}\xspace}                       
\def\Opep#1   {\ensuremath{\mathcal{O}_{#1}^{'}}\xspace}                    

\newcommand{\tev}{\ensuremath{\mathrm{\,Te\kern -0.1em V}}\xspace}
\newcommand{\gev}{\ensuremath{\mathrm{\,Ge\kern -0.1em V}}\xspace}
\newcommand{\mev}{\ensuremath{\mathrm{\,Me\kern -0.1em V}}\xspace}
\newcommand{\kev}{\ensuremath{\mathrm{\,ke\kern -0.1em V}}\xspace}
\newcommand{\ev}{\ensuremath{\mathrm{\,e\kern -0.1em V}}\xspace}
\newcommand{\gevc}{\ensuremath{{\mathrm{\,Ge\kern -0.1em V\!/}c}}\xspace}
\newcommand{\mevc}{\ensuremath{{\mathrm{\,Me\kern -0.1em V\!/}c}}\xspace}
\newcommand{\gevcc}{\ensuremath{{\mathrm{\,Ge\kern -0.1em V\!/}c^2}}\xspace}
\newcommand{\gevgevcccc}{\ensuremath{{\mathrm{\,Ge\kern -0.1em V^2\!/}c^4}}\xspace}
\newcommand{\mevcc}{\ensuremath{{\mathrm{\,Me\kern -0.1em V\!/}c^2}}\xspace}

\def\gsim{{~\raise.15em\hbox{$>$}\kern-.85em
          \lower.35em\hbox{$\sim$}~}\xspace}
\def\lsim{{~\raise.15em\hbox{$<$}\kern-.85em
          \lower.35em\hbox{$\sim$}~}\xspace}

\def\pt         {\mbox{$p_{\rm T}$}\xspace}

\def\evtgen     {\mbox{\textsc{EvtGen}}\xspace}

\def\geant      {\mbox{\textsc{Geant4}}\xspace}

\def\tell1  {TELL1\xspace}
\def\ukl1   {UKL1\xspace}

\newcommand{\ie}{\mbox{\itshape i.e.}}

\usepackage{mciteplus}

\begin{document}



\begin{titlepage}
\pagenumbering{roman}

\vspace*{-1.5cm}
\centerline{\large EUROPEAN ORGANIZATION FOR NUCLEAR RESEARCH (CERN)}
\vspace*{1.5cm}
\hspace*{-0.5cm}
\begin{tabular*}{\linewidth}{lc@{\extracolsep{\fill}}r}
\ifthenelse{\boolean{pdflatex}}
{\vspace*{-2.7cm}\mbox{\!\!\!\includegraphics[width=.14\textwidth]{figs/lhcb-logo.pdf}} & &}%
{\vspace*{-1.2cm}\mbox{\!\!\!\includegraphics[width=.12\textwidth]{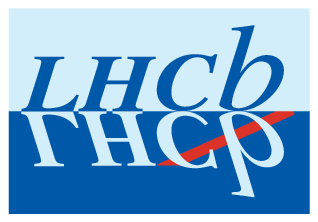}} & &}%
\\
 & & CERN-PH-EP-2012-171 \\  
 & & LHCb-PAPER-2011-037 \\  
 & & \today \\ 
 & & \\
\end{tabular*}

\vspace*{4.0cm}

{\bf\boldmath\huge
\begin{center}
  Measurement of prompt hadron production ratios in $pp$ collisions at $\sqrt{s} = $ 0.9 and 7 TeV
\end{center}
}

\vspace*{2.0cm}

\begin{center}
The LHCb collaboration\footnote{Authors are listed on the following pages.}
\end{center}

\vspace{\fill}

\begin{abstract}
  \noindent
The charged-particle  production ratios $\bar{p}/p$, $K^-/K^+$, $\pi^-/\pi^+$, $(p + \bar{p})/(\pi^+ + \pi^-)$, $(K^+ + K^-)/(\pi^+ + \pi^-)$ and $(p + \bar{p})/(K^+ + K^-)$ are measured with the LHCb detector using $0.3\,{\rm nb^{-1}}$ of $pp$ collisions delivered by the LHC at $\sqrt{s} = 0.9$~TeV and $1.8\,{\rm nb^{-1}}$  at $\sqrt{s} = 7$~TeV.  The measurements are performed as a function of transverse momentum \pt and pseudorapidity $\eta$. 
 The production ratios are compared to the predictions of several Monte Carlo generator settings, none of which are able to describe adequately all observables.  The ratio $\bar{p}/p$ is also considered as a function of rapidity loss, $\Delta y \equiv y_{\rm beam} - y$, and is used to constrain models of baryon transport. 
\end{abstract}

\begin{center}
Accepted by Eur. Phys. J. C
\end{center}

\vspace*{2.0cm}
\vspace{\fill}

\end{titlepage}


\newpage
\setcounter{page}{2}
\mbox{~}
\newpage

\centerline{\large\bf  LHCb collaboration}
\begin{flushleft}
\small
R.~Aaij$^{38}$, 
C.~Abellan~Beteta$^{33,n}$, 
A.~Adametz$^{11}$, 
B.~Adeva$^{34}$, 
M.~Adinolfi$^{43}$, 
C.~Adrover$^{6}$, 
A.~Affolder$^{49}$, 
Z.~Ajaltouni$^{5}$, 
J.~Albrecht$^{35}$, 
F.~Alessio$^{35}$
% ...

{\footnotesize \it
% ...
$ ^{5}$Clermont Universit\'{e}, Universit\'{e} Blaise Pascal, CNRS/IN2P3, LPC, Clermont-Ferrand, France\\
$ ^{6}$CPPM, Aix-Marseille Universit\'{e}, CNRS/IN2P3, Marseille, France\\
% ...
$ ^{11}$Physikalisches Institut, Ruprecht-Karls-Universit\"{a}t Heidelberg, Heidelberg, Germany\\
% ...
$ ^{33}$Universitat de Barcelona, Barcelona, Spain\\
$ ^{34}$Universidad de Santiago de Compostela, Santiago de Compostela, Spain\\
$ ^{35}$European Organization for Nuclear Research (CERN), Geneva, Switzerland\\
% ...
$ ^{38}$Nikhef National Institute for Subatomic Physics, Amsterdam, The Netherlands\\
% ...
$ ^{43}$H.H. Wills Physics Laboratory, University of Bristol, Bristol, United Kingdom\\
% ...
$ ^{49}$Oliver Lodge Laboratory, University of Liverpool, Liverpool, United Kingdom\\
% ...
$ ^{n}$LIFAELS, La Salle, Universitat Ramon Llull, Barcelona, Spain\\
% ...
}
\end{flushleft}

\cleardoublepage




\pagestyle{plain} 
\setcounter{page}{1}
\pagenumbering{arabic}


%

\section{Introduction}
\label{sec:intro}

All underlying interactions responsible for $pp$ collisions at the Large Hadron Collider (LHC) and the subsequent hadronisation process can be understood within the context of quantum chromodynamics (QCD). In the non-perturbative regime, however,  precise calculations are difficult to perform and so phenomenological models must be employed.  Event generators based on these models must be optimised, or `tuned', to reproduce experimental observables.  
The observables exploited for this purpose include event variables, such as particle multiplicities, the kinematical distributions of the inclusive particle sample in each event, and  the corresponding distributions for  individual particle species.
The generators can then be used in simulation studies when analysing data to search for physics beyond the Standard Model.  

The relative proportions of each charged quasi-stable hadron, and the ratio of antiparticles to particles in a given kinematical region, are important inputs for generator tuning.  Of these observables, the ratio of antiprotons to protons is of particular interest. 
Baryon number conservation requires that the disintegration of the beam particles that occurs in high-energy inelastic non-diffractive
$pp$ collisions must be balanced by the creation of protons or other baryons elsewhere in the event.  This topic is
known as {\it baryon-number transport}.  Several models exist to describe this transport, but it is not clear  which mechanisms are most important in driving the phenomenon\cite{THEORY1,THEORY2,THEORY3a,THEORY3b,THEORY3c,THEORY4a,THEORY4b,THEORY4c,THEORY5,THEORY6a,THEORY6b,THEORY7,THEORY8}.  
Pomeron exchange is expected to play a significant role, but contributions may exist from other sources, for example the Odderon, the existence of which has not yet been established\cite{THEORY8,ODDERON1,ODDERON2}.
Experimentally, baryon-number transport can be studied by measuring $\bar{ p}/{ p}$, the ratio of the number of produced antiprotons to protons, as a function of suitable kinematical variables.

In this paper results are presented from the LHCb experiment for the following production ratios:  $\bar{ p}/{ p}$, $K^-/K^+$, $\pi^-/\pi^+$,  
$(p + \bar{p})/(\pi^+ + \pi^-)$, $(K^+ + K^-)/(\pi^+ + \pi^-)$ and $(p + \bar{p})/(K^+ + K^-)$.    The first three of these observables are termed the {\it same-particle ratios} and the last three the {\it different-particle ratios}. Only prompt particles are considered, 
where a prompt particle is defined to be one that originates from the primary interaction, either directly, or through the subsequent decay of a resonance.
The ratios are measured  as a function of transverse momentum \pt and pseudorapidity $\eta = -\ln (\tan \theta/2)$, where $\theta$ is the polar angle with respect to the beam axis.

Measurements have been performed of the $\bar{p}/p$ ratio in $pp$ collisions both at the LHC\cite{ALICEPBARP}, and at other facilities\cite{ISR,BRAHMS, PHENIX,PHOBOS,STAR,NA49}.  Studies have also been made of the production characteristics of pions, kaons and protons at the LHC at $\sqrt{s} =0.9$~TeV at mid-rapidity\cite{ALICERATIOS}.  The analysis presented in this paper exploits the unique forward coverage of the LHCb spectrometer, and the powerful particle separation capabilities of the ring-imaging Cherenkov (RICH) system, to yield results for the production ratios in the range $2.5 < \eta < 4.5$ at both $\sqrt{s}=0.9$~TeV and  $\sqrt{s}=7$~TeV.  LHCb has previously published studies of baryon transport and particle ratios with neutral strange hadrons\cite{LHCBV0}, and results for strange baryon observables at the LHC are also available in the midrapidity region\cite{ATLASV0,ALICESTRANGE}. New analyses have also been made public  since the submission of this paper~\cite{CMSNEW}.

The paper is organised as follows.  Section~\ref{sec:dataset} introduces the LHCb detector and the datasets used. Section~\ref{sec:selection}  describes the selection of the analysis sample, while Sect.~\ref{sec:pid} discusses the calibration of the particle identification performance. The analysis procedure is  explained in Sect.~\ref{sec:analysis}.  The assignment of the systematic uncertainties is described in Sect.~\ref{sec:systs} and the results are presented and discussed in Sect.~\ref{sec:results}, before concluding in Sect.~\ref{sec:conclusions}.
Full tables of numerical results may be found in Appendix~\ref{sec:tabresults}.  Throughout, unless specified otherwise, particle types are referred to by their name ({\it e.g.} proton) when both particles and antiparticles are being considered together, and by symbol ({\it e.g.} $p$ or $\bar{p}$) when it is necessary to distinguish between the two.

\section{Data samples and the LHCb detector}
\label{sec:dataset}

The LHCb experiment is a forward spectrometer at the Large Hadron Collider with a pseudorapidity
acceptance of approximately $2 < \eta < 5$.   The tracking system
begins with a silicon strip Vertex Locator (VELO). The VELO consists 
of 23 sequential stations
of silicon strip detectors which retract from the beam during injection.
A large area silicon tracker (TT) follows upstream of a dipole magnet,
downstream of which there are three tracker stations, each built with a mixture of 
straw tube and silicon strip detectors.  
The dipole field direction is vertical, and charged tracks reconstructed through 
the full spectrometer are deflected by an integrated $B$ field of around 4~Tm.
Hadron identification is provided by the RICH system, which
consists of two detectors, one upstream
of the magnet and the other downstream, and is designed to provide
particle identification over a momentum interval of 2--100~GeV/$c$.
Also present, but not exploited in the current analysis, are a calorimeter and muon system.
A full description of the LHCb detector 
may be found in~\cite{LHCb}. 

The data sample under consideration derives from the early period 
of the 2010 LHC run.  Inelastic interactions were triggered 
by requiring at least one track in either the VELO
or the tracking stations downstream of the magnet.  
This trigger was more than 99\% efficient for all offline selected events that contain at least two tracks reconstructed 
through the whole system.
Collisions
were recorded both at $\sqrt{s}=0.9$~TeV and $7$~TeV. 
During $0.9$~TeV running,
where the beams were wider and the internal crossing-angle of the beams
within LHCb was larger, detector and machine safety considerations required
that each VELO half was retracted by 10~mm from the nominal closed position.
For 7~TeV operation the VELO was fully closed.

The analysis exploits a data sample of around $0.3\,{\rm nb}^{-1}$ recorded
at $\sqrt{s}= 0.9$~TeV and $1.8 \,{\rm nb}^{-1}$ at $\sqrt{s}=7$~TeV.
In order to minimise  potential detector-related 
systematic biases, the direction of the LHCb dipole field was
inverted every 1--2 weeks of data taking. At 0.9~TeV the data
divide approximately equally between the two polarities, while at 7~TeV around two-thirds were collected in one configuration. 
The analysis is performed separately for each polarity.

The beams collided with a crossing angle in the horizontal plane which was set to 
compensate for the field of the LHCb dipole.  This angle was 2.1~mrad in magnitude at $\sqrt{s} = 0.9$~TeV 
and 270~$\upmu$rad at  $\sqrt{s} = 7$~TeV.  Throughout this analysis momenta and any derived quantities are computed in the centre-of-mass frame.

Monte Carlo simulated events are used to calculate efficiencies
and estimate systematic uncertainties. A total of around 140 million events are simulated 
at 0.9~TeV and 130 million events at 7~TeV. 
The $pp$ collisions are generated by \textsc{Pythia6.4}~\cite{PYTHIA} and the parameters tuned as described in  Ref.~\cite{VANYA}.
The decays of emerging particles are implemented  
with the \evtgen package~\cite{EVTGEN}, with final state radiation described by \textsc{Photos}~\cite{PHOTOS}.
The  resulting particles are transported through LHCb by \geant~\cite{GEANT,GEANTXTRA}, which models hits
in the sensitive regions of the detector as well as material interactions as described in Ref.~\cite{CLEMENCIC}.  The decay of secondary particles produced in
these interactions is controlled by \geant.  Additional \textsc{Pythia6.4} samples with different generator tunes were produced 
in order to provide references with which to compare the results.  These were Perugia~0, which was tuned on
experimental results from SPS, LEP and the Tevatron,  and Perugia~NOCR, which includes an extreme model of baryon transport\cite{PERUGIA}.

\section{Selection of the analysis sample}
\label{sec:selection}

The measurement is performed using the {\it analysis sample}, the selection of which is described here. Understanding of the particle identification (PID) performance provided by the RICH sample  is obtained from the {\it calibration sample}, which is discussed in  Sect.~\ref{sec:pid}.

Events are selected which contain at least one reconstructed primary vertex (PV) within
$20$~cm of the nominal interaction point.    The primary vertex finding algorithm requires at least 
three reconstructed tracks.\footnote{The PV requirement can be approximated in Monte Carlo simulation by imposing a filter at generator level which demands at least three charged particles with lifetime $c\tau > 10^{-9}$~m, momentum $p > 0.3$~GeV/$c$ and polar angle $15 < \theta < 460$~mrad.}

Tracks are only considered 
that have hits both in the VELO detector
and in the tracking stations downstream of the magnet, and for which the track fit yields an acceptable  $\chi^2$ per 
number of degrees of freedom (ndf).
In order to suppress background from  decays of long-lived particles, or particles produced in secondary interactions, an upper bound is placed on the goodness of fit when using the track's impact parameter (IP) to test the hypothesis that the track is associated 
with the PV ($\chi^2_{\rm IP} < 49$).  To reduce systematic uncertainties in the calculation of the ratio observables, a momentum cut is imposed of $p> 5$~GeV/$c$, as below this value the cross-section for strong interaction with the beampipe and detector elements differs significantly between particle and anti-particle for kaons and protons.
If a pair of tracks, $i$ and $j$, are found to have very similar momenta ($|{\mathbf p}_i -{\mathbf  p}_j|/|{\mathbf p}_i +{\mathbf p}_j| < 0.001$), then one of the two is rejected at random.  This requirement is imposed to suppress `clones', which occur when two tracks are reconstructed from the hit points left by a single particle, and eliminates ${\cal{O}}(1\%)$ of candidates.


The analysis is performed in bins of \pt and $\eta$.  In $p_{\rm T}$ three
separate regions are considered: $p_{\rm T} < 0.8$~GeV/$c$, 
$0.8 \le p_{\rm T} < 1.2$~GeV/$c$ and $p_{\rm T} \ge 1.2$~GeV/$c$.  In $\eta$ half-integer bins are chosen over the intervals  $3.0 < \eta < 4.5$  for $p_{\rm T} < 0.8$~GeV/$c$,   and $2.5 <  \eta < 4.5$ for higher $p_{\rm T}$  values. The  $\eta$ acceptance is not constant with $p_{\rm T}$  because the limited size of the calibration samples does not allow for the PID performance to be determined with adequate precision below $\eta =3$ in the lowest $p_{\rm T}$ bin.   The bin size is large compared to the experimental resolution and hence bin-to-bin migration effects are negligible in the analysis.

\begin{figure}
\begin{center}
\includegraphics[width=0.48\textwidth]{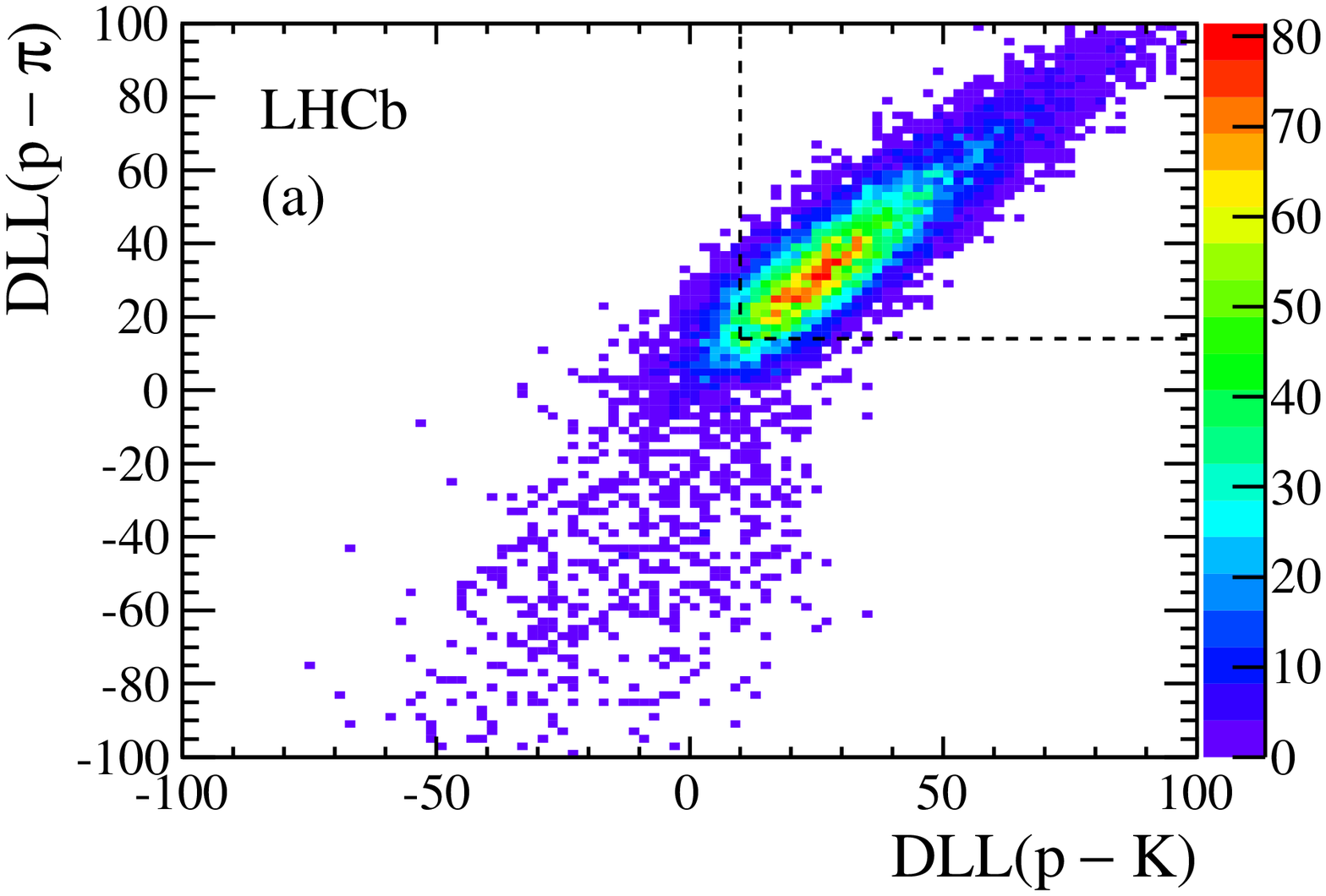}
\includegraphics[width=0.48\textwidth]{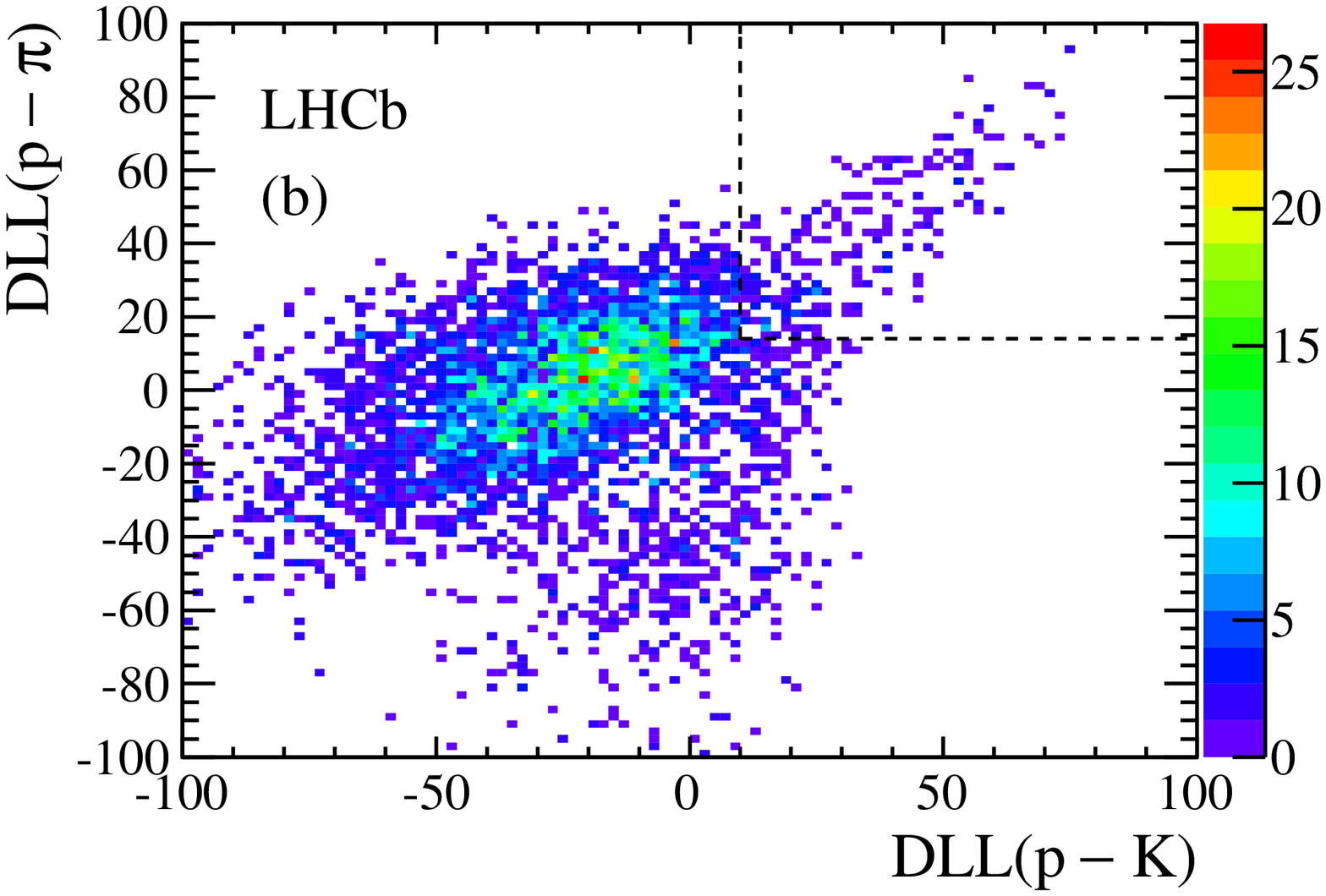}
\includegraphics[width=0.48\textwidth]{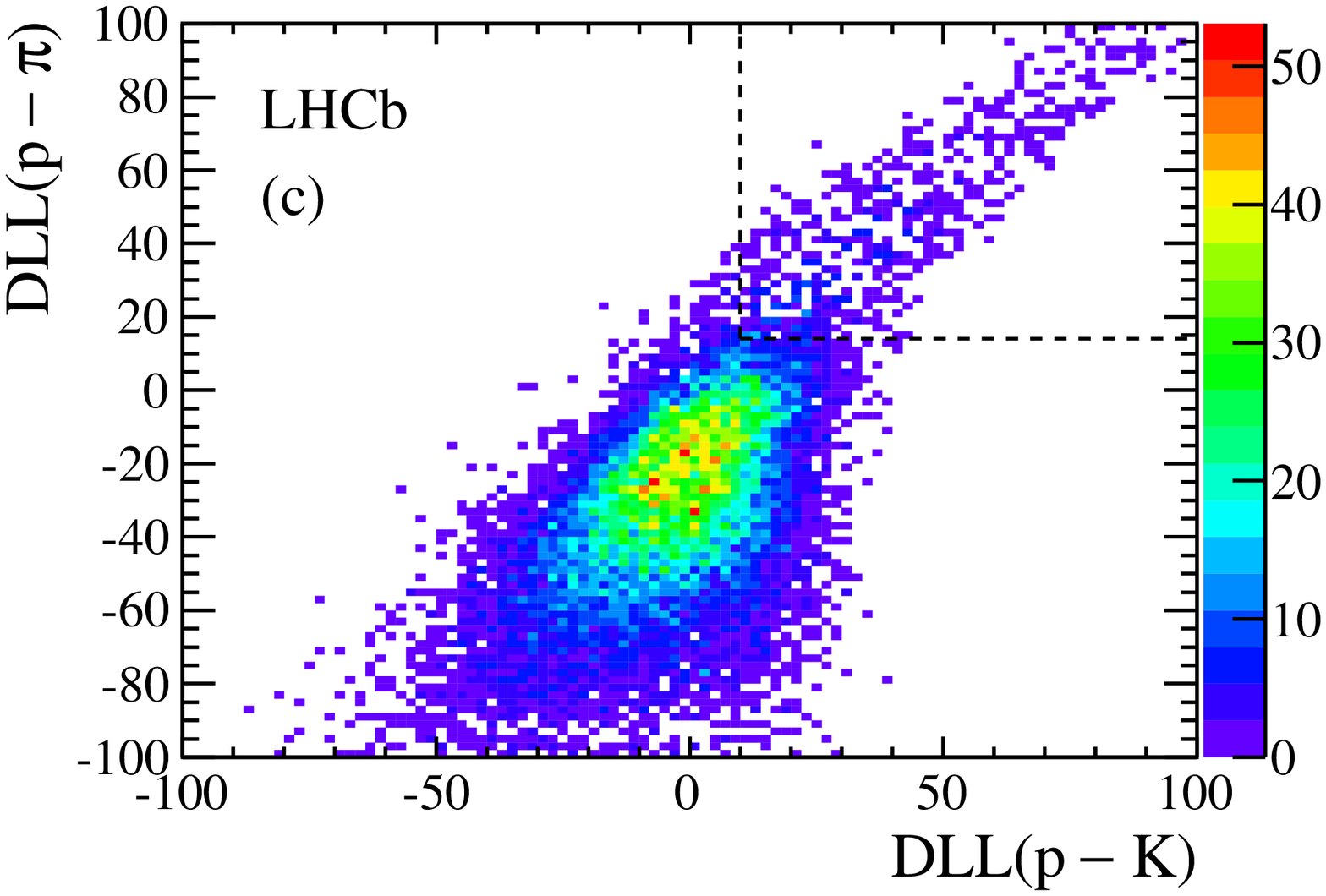}
\caption{\small Two-dimensional distribution of the change in log likelihood DLL($p-K$) and DLL($p-\pi$) for (a) protons, (b) kaons and (c) pions (here shown 
for negative tracks and one magnet polarity)
in the calibration sample  with $p_{\rm T} > 1.2$~GeV/$c$
and $3.5 < \eta \le 4.0$.  The region indicated by the dotted lines
in the top right corner of each plot is that which is selected in
the analysis to isolate the proton sample.  The selection of the calibration sample is discussed in Sect.~\ref{sec:pid}.}
\label{fig:LLpid2D}
\end{center}
\end{figure}

The RICH is used to select the analysis sample at both energy points from which the ratio observables are determined.
A  pattern recognition
and particle identification algorithm uses information from the RICH and tracking detectors to construct
a negative log likelihood for each particle hypothesis  ($e$, $ \mu$, $ \pi$, $K$ or $p$).
This negative log likelihood is minimised
for the event as a whole. After minimisation, the change
in log likelihood (DLL) is recorded for each track when the particle type
is switched from that of the preferred assignment to another hypothesis.
Using this information the separation in log likelihood DLL($x-y$) can be calculated for any two particle hypotheses $x$ and $y$, 
where a positive value indicates that $x$ is the favoured option.
In the analysis, cuts are placed on DLL($ p-K$) versus DLL($ p-\pi$) 
to select protons and on DLL($ K-p$) versus DLL($ K-\pi$) to select kaons.
Pions are selected with a simple cut on DLL($ \pi-K$).  As the RICH performance
varies with momentum and track density,  different cuts are applied in each $(p_{\rm T}, \eta)$ bin.
The selection cuts are chosen in 
order to optimise purity, together with the requirement that the identification efficiency be at least 10\%.
Figure~\ref{fig:LLpid2D} shows the background-subtracted two-dimensional distribution of  DLL($ p-K$) and DLL($ p-\pi$) 
for protons, kaons and pions in the calibration sample for one example bin.
The approximate number of positive and negative tracks selected  in each PID category is given in Tables~\ref{tab:sample_low} and~\ref{tab:sample_high}.~\footnote{The journal version of this paper has incorrect entries in Table~\ref{tab:sample_high}.}  A charge asymmetry can be observed in many bins, most noticeably for the protons.

\begin{table}
\begin{center}
\caption{\small Number of particle candidates in the analysis sample at $\sqrt{s}=0.9$ TeV, separated into positive and negative charge ($Q$).}~\label{tab:sample_low}
\begin{tabular}{l|r|rrr|rrr|rrr}
   & & \multicolumn{3}{c|}{$p_{\rm T} < 0.8$ GeV/$c$} & \multicolumn{3}{c|}{$0.8 \le p_{\rm T} < 1.2$ GeV/$c$ } & \multicolumn{3}{c}{$p_{\rm T} \ge 1.2$ GeV/$c$} \\ 
  & $Q$  & \multicolumn{1}{c}{\hspace*{0.60cm}$p$\hspace*{0.0cm}} & \multicolumn{1}{c}{\hspace*{0.60cm}$K$\hspace*{0.0cm}} &
\multicolumn{1}{c|}{\hspace*{0.62cm}$\pi$\hspace*{0.0cm}} & \multicolumn{1}{c}{\hspace*{0.62cm}$p$\hspace*{0.0cm}} & \multicolumn{1}{c}{\hspace*{0.62cm}$K$\hspace*{0.0cm}} & \multicolumn{1}{c|}{\hspace*{0.62cm}$\pi$\hspace*{0.0cm}} & \multicolumn{1}{c}{\hspace*{0.62cm}$p$\hspace*{0.0cm}} & \multicolumn{1}{c}{\hspace*{0.62cm}$K$\hspace*{0.0cm}}
 & \multicolumn{1}{c}{\hspace*{0.62cm} $\pi$ \hspace*{0.0cm}}  \\ \hline
$2.5 < \eta <  3.0$ & $+$ & -- & -- & -- & 16k & 39k & 270k & 19k & 36k & 130k \\
 & $-$  & -- & -- & -- & 13k & 35k & 270k & 13k & 31k & 120k \\
&&&&&&&&&&\\
$3.0 \le \eta < 3.5 $ & $+$ &21k & 78k & 1.1M & 30k & 63k & 260k & 34k & 39k & 120k \\
 & $-$  & 17k & 69k & 1.1M & 21k & 55k & 250k & 20k & 31k & 100k \\
&&&&&&&&&&\\
$3.5 \le \eta < 4.0$ & $+$ &  55k & 120k & 1.9M & 55k & 60k & 240k & 31k & 33k & 97k \\
 & $-$ &  38k & 100k & 1.9M & 33k & 49k & 230k &  14k & 23k & 85k \\
&&&&&&&&&&\\
$4.0 \le \eta <  4.5$ & $+$ &  26k & 90k & 1.2M &  23k & 30k & 100k & 14k & 11k & 39k \\
 &$ -$  &  21k & 86k & 1.2M & 11k & 22k & 88k &  4.2k & 6.6k & 30k \\
\end{tabular}
\end{center}
\end{table}

\begin{table}
\begin{center}
\caption{\small Number of particle candidates  in the analysis sample at $\sqrt{s}=7.0$ TeV, separated into positive and negative charge ($Q$).}~\label{tab:sample_high}
\begin{tabular}{l|r|rrr|rrr|rrr}
 & & \multicolumn{3}{c|}{$ p_{\rm T} < 0.8$ GeV/$c$} & \multicolumn{3}{c|}{$0.8 \le p_{\rm T} < 1.2$ GeV/$c$ } & \multicolumn{3}{c}{$p_{\rm T} \ge 1.2$ GeV/$c$} \\ 
 & $Q$ & \multicolumn{1}{c}{\hspace*{0.60cm}$p$\hspace*{0.0cm}} & \multicolumn{1}{c}{\hspace*{0.60cm}$K$\hspace*{0.0cm}} & \multicolumn{1}{c|}{\hspace*{0.62cm}$\pi$\hspace*{0.0cm}} & \multicolumn{1}{c}{\hspace*{0.62cm}$p$\hspace*{0.0cm}} & \multicolumn{1}{c}{\hspace*{0.62cm}$K$\hspace*{0.0cm}} & \multicolumn{1}{c|}{\hspace*{0.62cm}$\pi$\hspace*{0.0cm}} & \multicolumn{1}{c}{\hspace*{0.62cm}$p$\hspace*{0.0cm}} & \multicolumn{1}{c}{\hspace*{0.62cm}$K$\hspace*{0.0cm}}
 & \multicolumn{1}{c}{\hspace*{0.62cm} $\pi$ \hspace*{0.0cm}}  \\ \hline
$2.5 < \eta <  3.0$ & $+$ & -- & -- & -- & 180k & 850k & 6.8M & 500k & 1.3M & 4.6M \\
 & $-$  & -- & -- & -- & 170k & 820k & 6.8M &  450k & 1.2M & 4.7M \\
&&&&&&&&&&\\
$3.0 \le \eta < 3.5 $ & $+$ & 230k & 1.5M & 22M &  380k & 1.6M & 6.7M & 850k & 1.4M & 4.4M \\
 & $-$  &  220k & 1.4M & 23M & 350k & 1.6M & 6.7M &  760k & 1.4M & 4.4M \\
&&&&&&&&&&\\
$3.5 \le \eta < 4.0$ & $+$ &  740k & 2.5M & 38M & 930k & 1.6M & 6.4M & 880k & 1.2M & 3.8M \\
 & $-$ &  690k & 2.4M & 38M & 840k & 1.5M & 6.3M & 760k & 1.2M & 3.7M \\
&&&&&&&&&&\\
$4.0 \le \eta <  4.5$ & $+$ &  460k & 3.4M & 44M & 490k & 1.3M & 4.6M & 480k & 650k & 2.6M \\
 &$ -$  & 450k & 3.2M & 43M & 420k & 1.3M & 4.4M & 390k & 580k & 2.5M \\
\end{tabular}
\end{center}
\end{table}

\vspace{-0.3cm}
\section{Calibration of particle identification}
\label{sec:pid}

The calibration sample consists of the decays\footnote{In this section the inclusion of the charge conjugate decay $\bar{\Lambda} \to \bar{p} \pi^+ $ is implicit.} $ K^0_{\rm S} \to \pi^+\pi^-$, $ \Lambda \to p \pi^- $ and $ \phi \to K^+K^- $, all selected from the 7~TeV data.  The signal yields in each category are 4.7 million, 1.4 million and 5.5 million, respectively.
 
\begin{figure}[htb]
\begin{center}
\includegraphics[width=0.48\textwidth]{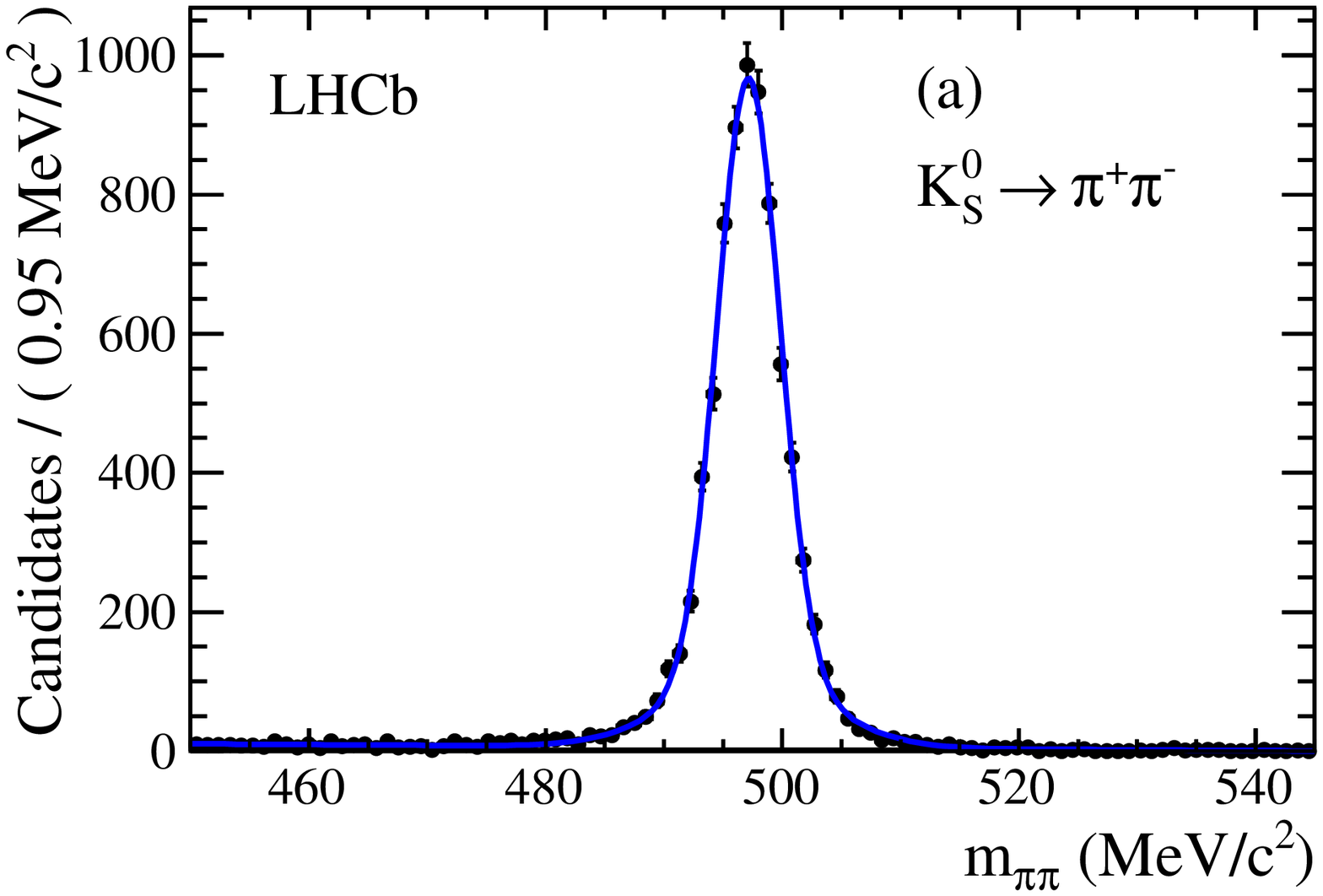}
%
\includegraphics[width=0.48\textwidth]{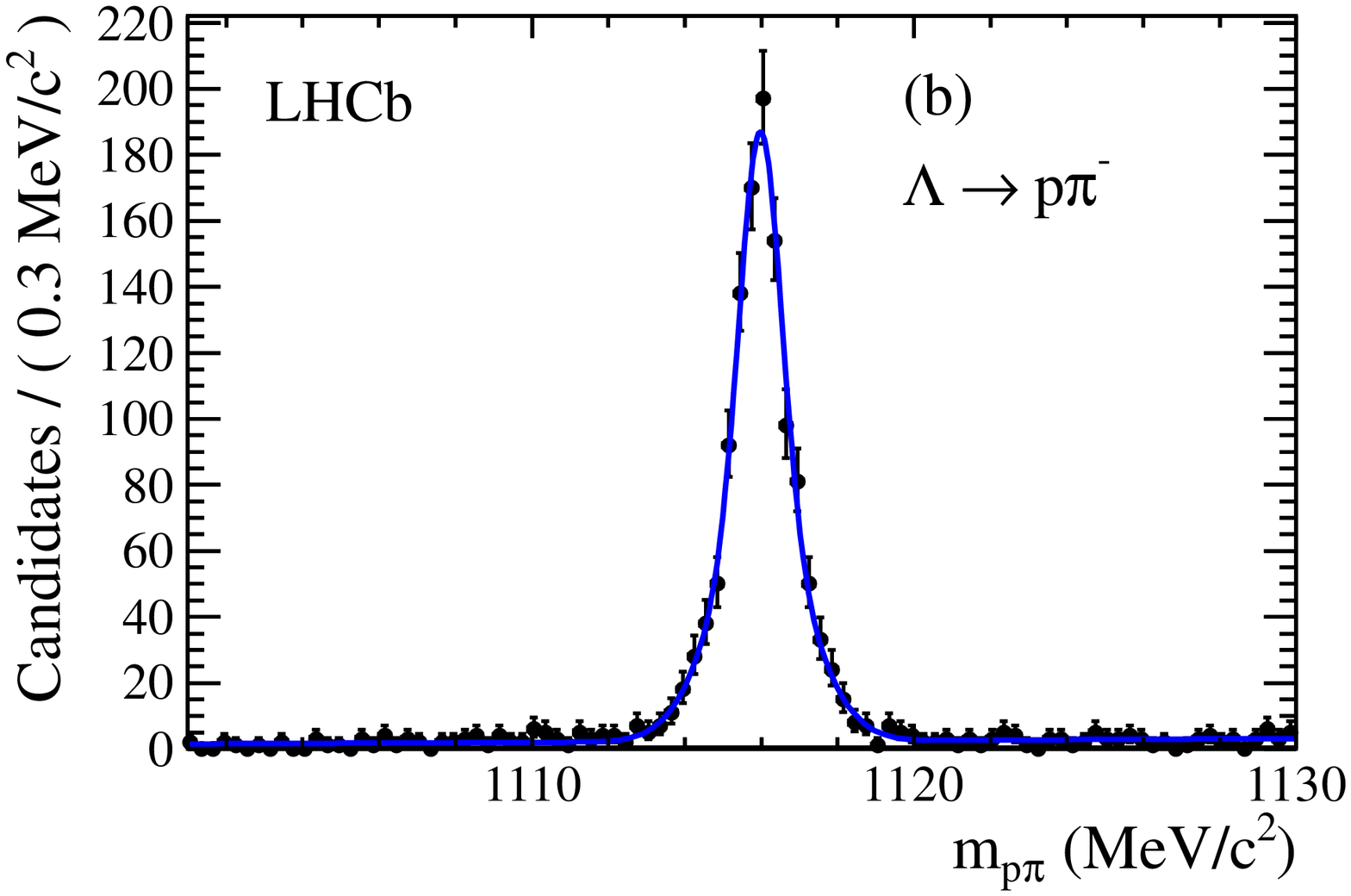}
\includegraphics[width=0.48\textwidth]{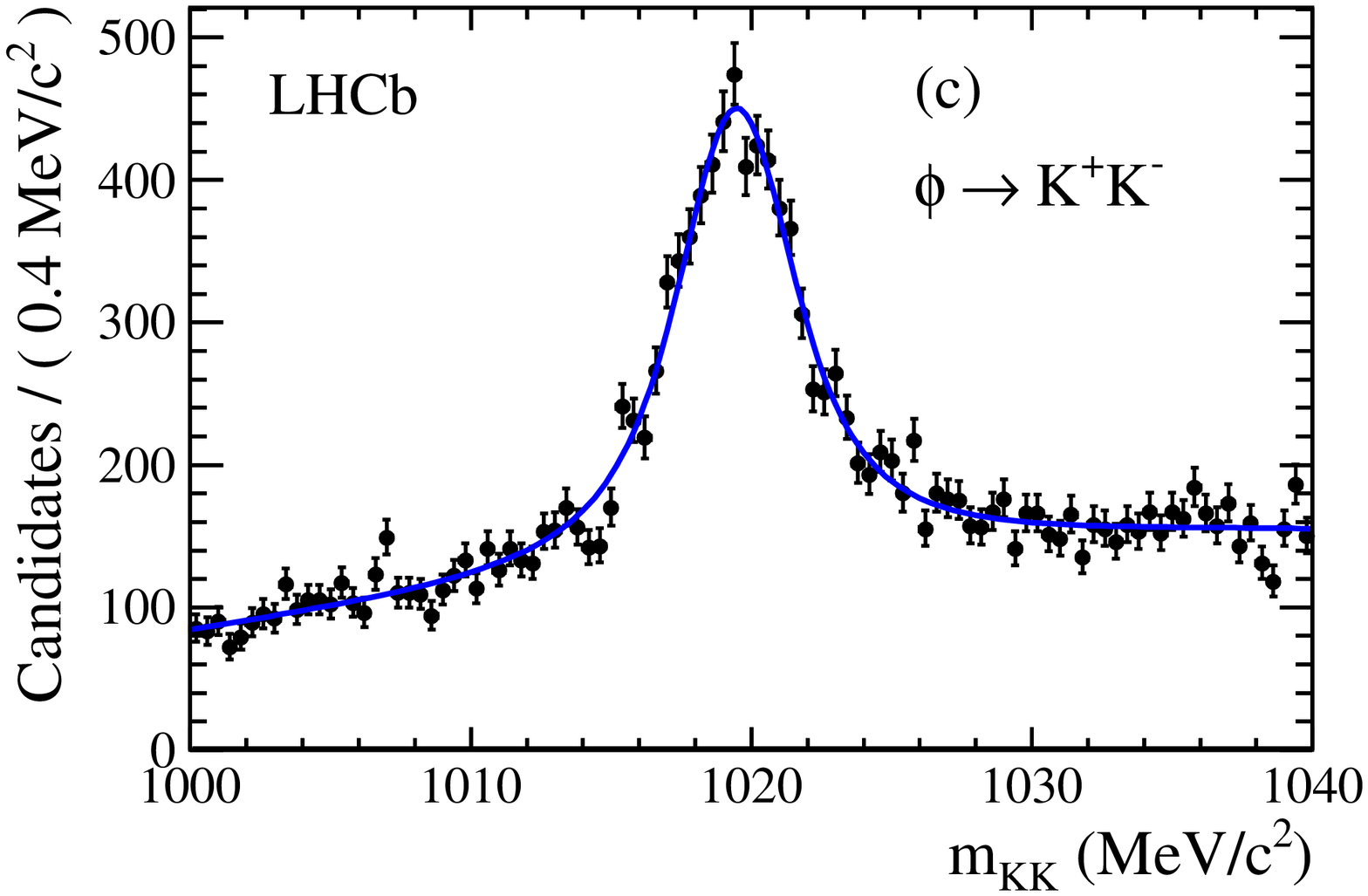}
\caption{\small Invariant mass distributions reconstructed for one magnet polarity from the $\sqrt{s}=7 \,\mathrm{TeV}$ data in the analysis bin for which the positive final-state particle has $\pt \ge 1.2~{\rm GeV}$ and $3.5\le \eta<4.0$ for (a) $K^0_{\rm S} \to \pi^+\pi^-$, (b)  $\Lambda \to p \pi^-$ and (c) $\phi \to K^+K^-$. The results of unbinned maximum likelihood fits to the data are superimposed.}
\label{fig:calib_samples}
\end{center}
\end{figure}

The $K^0_{\rm S}$ and $\Lambda$  (collectively termed $V^0$) decays are reconstructed through a selection algorithm devoid of RICH PID requirements, identical to that used in Ref.~\cite{LHCBV0}, providing samples of pions and protons which are unbiased for PID studies. 
The purity of the samples varies across the \pt and $\eta$ bins, but is found always to be in excess of 83\% and 87\%, for $K^0_{\rm S}$ and $\Lambda$, respectively.
Isolating $\phi \to K^+K^-$ decays with adequate purity is only achievable by exploiting RICH information. A PID requirement of DLL$(K -\pi)> 15$  is placed on one of the two kaon candidates, chosen at random,  so as to leave the other candidate  unbiased for calibration studies.  The purity of this selection ranges from 17\% to 68\%, over the kinematic range.
Examples of the invariant mass distributions obtained in a typical analysis bin for each of the three calibration modes are shown in Fig.~\ref{fig:calib_samples}.

In order to study the PID performance on the unbiased $K^{\pm}$ tracks associated with genuine $\phi$ decays
the {\it sPlot}\cite{SWEIGHTS} technique is employed, using the invariant mass as the uncorrelated discriminating variable,
to produce distributions of quantities such as the RICH DLL$(K -\pi)$.
Although the background contamination in the $V^{0}$ selections is small in comparison, the same strategy is employed to extract the true DLL distributions from all unbiased track samples in each analysis bin. 
The two $V^{0}$ signal peaks are parameterised by a double Gaussian function, while the strongly decaying $\phi$ is described by a 
Breit-Wigner function convoluted with a Gaussian.
The background is modelled by a first and third order Chebyshev polynomial for the $V^{0}$ and $\phi$ distributions, respectively.
 
The resulting distributions cannot be applied directly to the analysis sample for two reasons.
 The first is that the PID performance varies with momentum, and the finite size of the (\pt, $\eta$) bins means that the momentum spectrum within each bin is in general different between the calibration and analysis samples. The second is that the PID performance is also dependent on multiplicity, and here significant differences exist between the calibration and analysis samples, most noticeably for the 0.9~TeV data. To obtain rates applicable to the 0.9~TeV and 7~TeV analysis samples, it is therefore necessary to reweight the calibration tracks such that their distributions in momentum and track multiplicity match those of  a suitable reference sample.
A single  reference sample cannot be adopted for all particle types, as the unbiased momentum spectrum is in general different particle-to-particle.  For this reason, the analysis samples are used, but with the final selection replaced by looser PID requirements. This modified selection minimises distortions to the momentum spectra, while providing sufficient purity for the differences in distributions between particle species to be still evident.  In each (\pt, $\eta$) bin the reference and calibration samples are subdivided into six momentum and four track multiplicity cells, and the relative proportion of tracks within each cell is used to calculate a weight.  The PID performance as determined from the calibration samples after reweighting is then applied in the analysis.

\begin{figure}[htb]
\begin{center}
\includegraphics[width=0.48\textwidth]{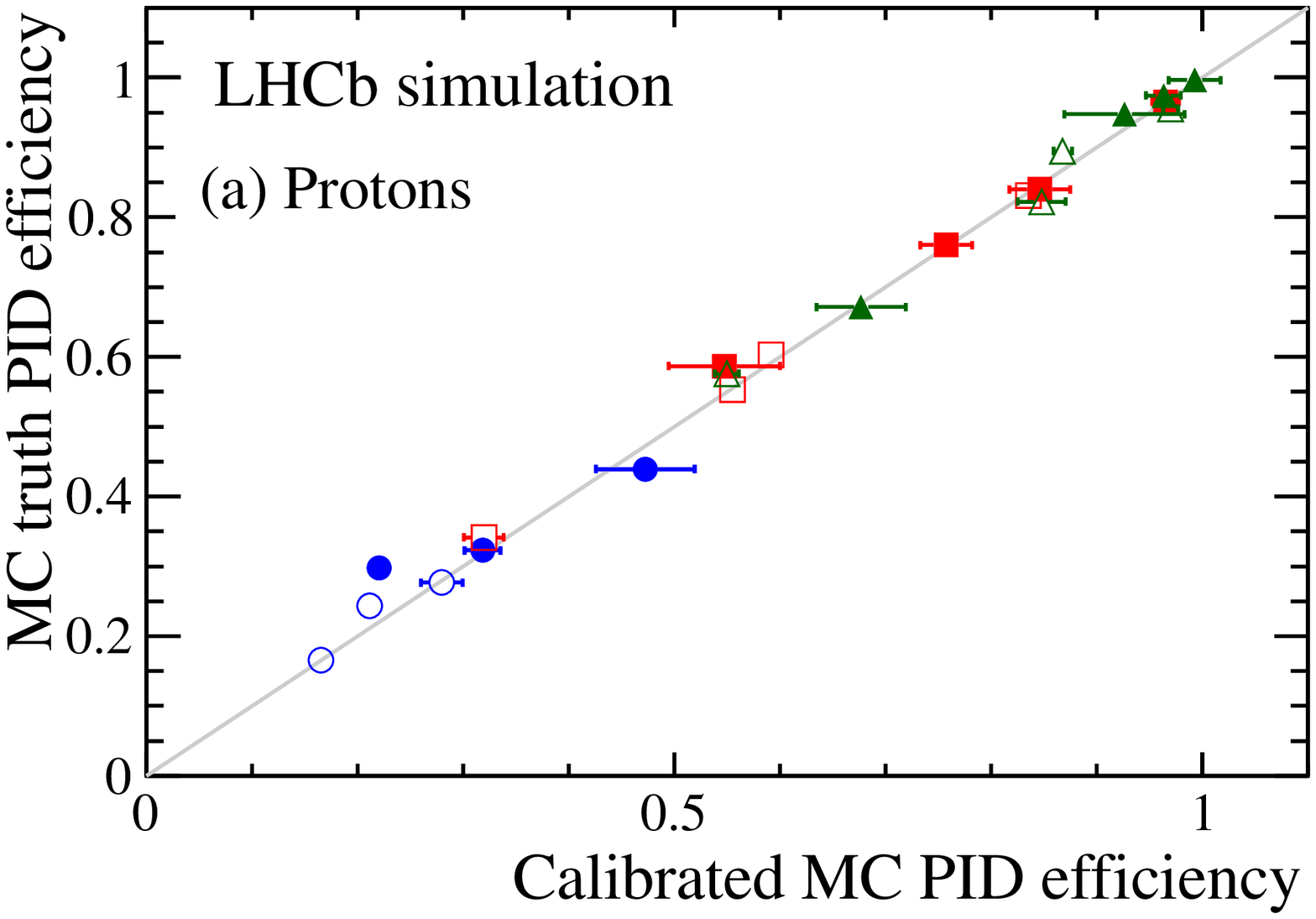}
\includegraphics[width=0.48\textwidth]{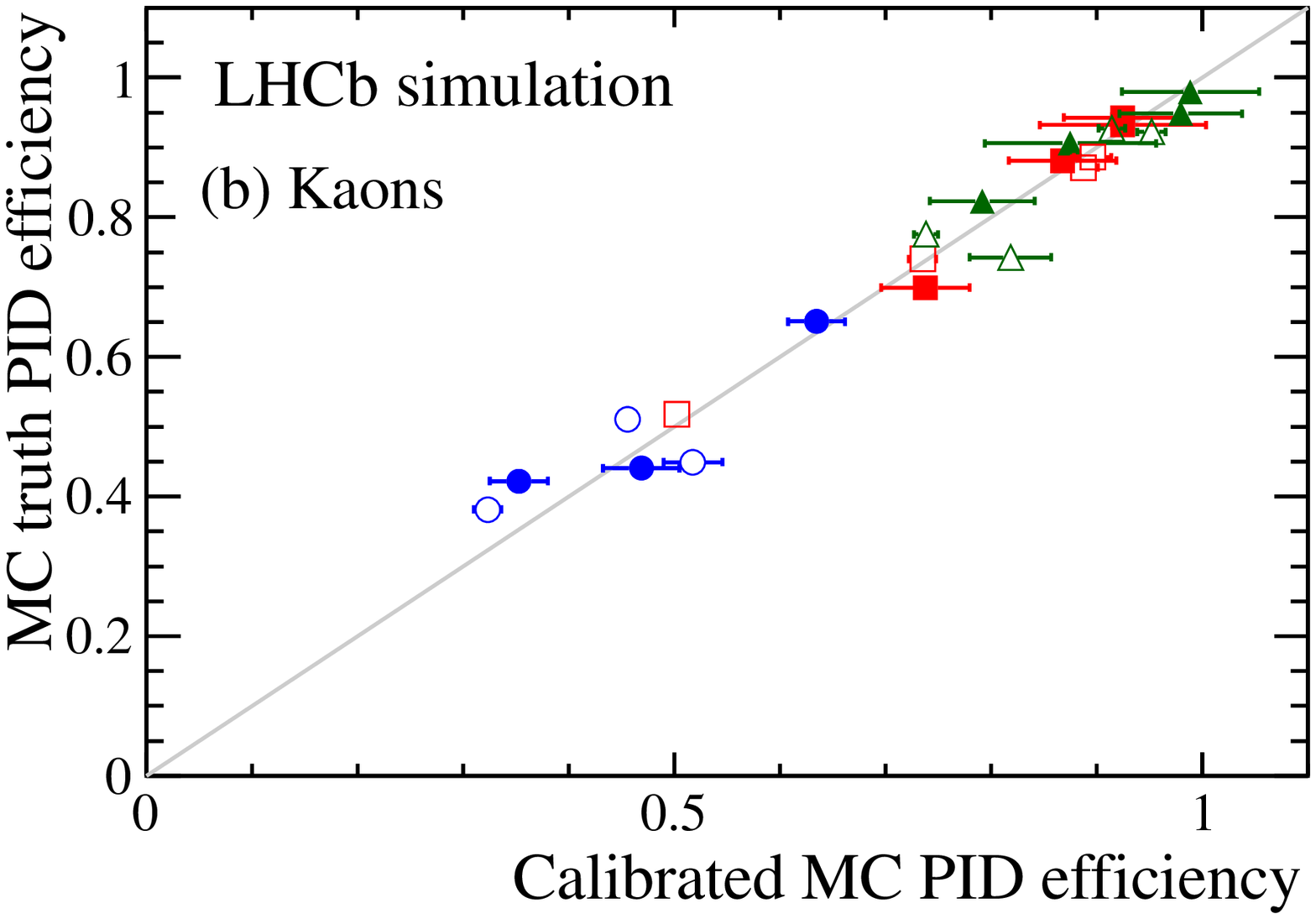}
\includegraphics[width=0.48\textwidth]{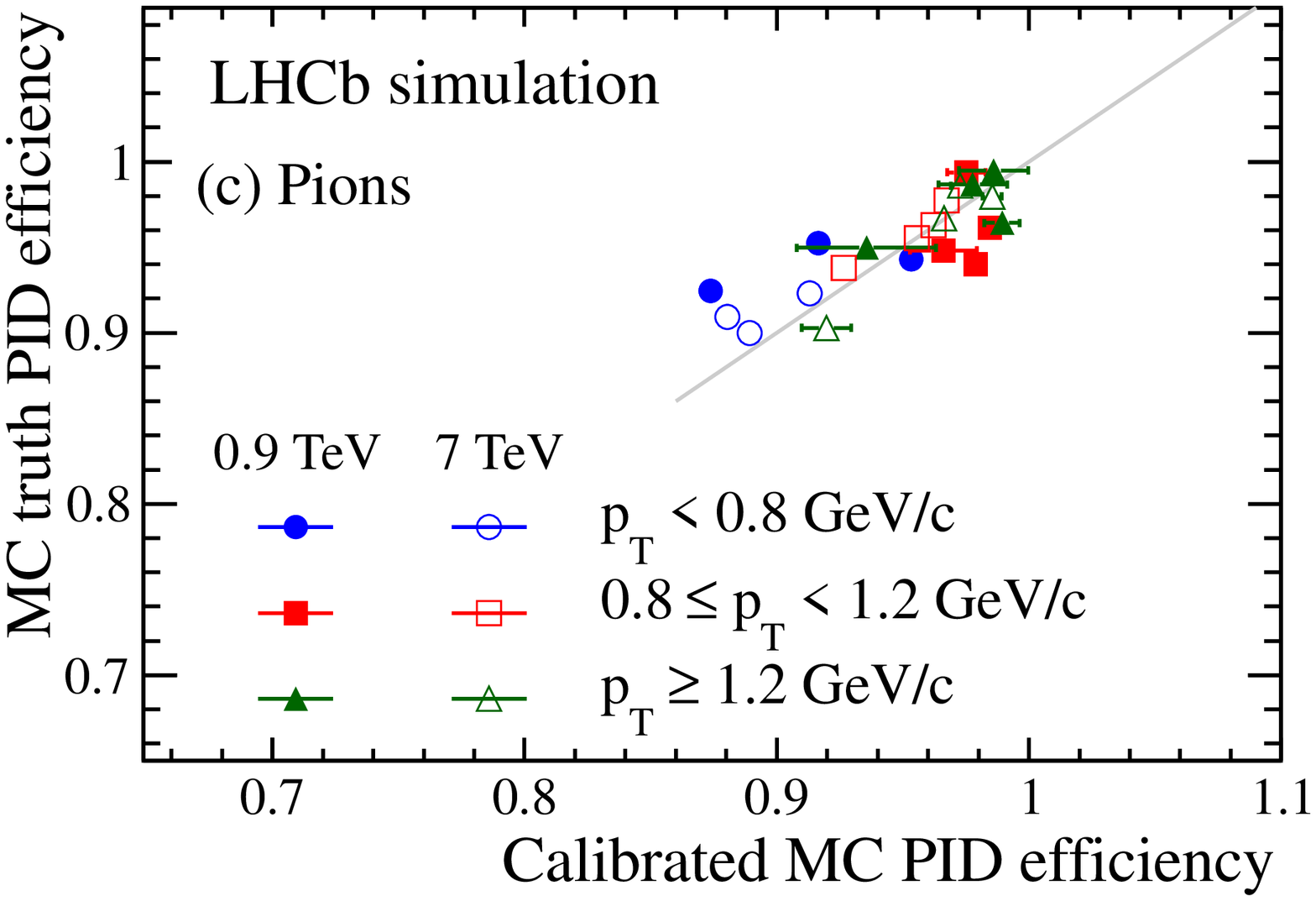}
\caption{\small Monte Carlo PID efficiency study for protons (a), kaons (b) and pions (c). Shown is a comparison of measured efficiencies from a Monte Carlo calibration sample, after background subtraction and reweighting, with the true values in the Monte Carlo analysis sample.
The diagonal line on each plot is drawn with unit gradient.
}
 \label{fig:mcpid}
\end{center}
\end{figure}

The reliability of the calibration  can be assessed by comparing the results for the measured PID efficiencies from a Monte Carlo simulated calibration sample, after background subtraction and reweighting, to the true values in the Monte Carlo analysis sample.  The results are shown in Fig.~\ref{fig:mcpid}, where each entry comes from a separate (\pt, $\eta$) bin. In general good agreement is observed over a wide range of working points, with some residual biases seen at low \pt.  These biases can be attributed to minor deficiencies in the reweighting procedure, which are expected to be most prevalent in this region.

\section{Analysis procedure}
\label{sec:analysis}

The number of particles, $N^{\rm S}_i$, selected in each of the three classes $i=p, \, K$ or $\pi$, is related to the true number of particles before particle identification, $N^{\rm T}_i$, by the relationship
\begin{equation}
\left (
\begin{array}{c}
N^{\rm S}_p \\
N^{\rm S}_K \\
N^{\rm S}_\pi
\end{array}
\right )
\, = \,
\left (
\begin{array}{ccc}
\epsilon_{p \to p} & \epsilon_{K \to p} & \epsilon_{\pi \to p} \\
\epsilon_{p \to K} & \epsilon_{K \to K} & \epsilon_{\pi \to K} \\
\epsilon_{p \to \pi} & \epsilon_{K \to \pi} & \epsilon_{\pi \to \pi}
\end{array}
\right )
\left (
\begin{array}{c}
N^{\rm T}_p \\
N^{\rm T}_K \\
N^{\rm T}_\pi
\end{array}
\right ),
\label{eq:id_matrix}
\end{equation}
where the matrix element $\epsilon_{i \to j}$ is the probability of identifying particle type $i$ as type $j$. This expression is valid for the purposes of the measurement since the fraction of other particle types, in particular electrons and muons, contaminating the selected sample is negligible.   As $N^{\rm S}_i$ and $\epsilon_{i \to j}$ are known,  the expression can be inverted to determine $N^{\rm T}_i$.  This is done for each $(\pt, \eta)$ bin, at each energy  point and magnet polarity setting.  After this step (and including the low \pt scaling factor correction discussed below) the purities of each sample can be calculated.  Averaged over the analysis bins the purities at 0.9~TeV (7~TeV) are found to be 0.90 (0.84), 0.89 (0.87) and 0.98 (0.97) for the protons, kaons and pions, respectively. 

In order to relate $N^{\rm T}_i$ to the number of particles produced in the primary interaction
it is necessary to correct for the effects of  non-prompt contamination, geometrical acceptance losses and track finding inefficiency.  
The non-prompt correction, according to simulation, is typically 1--2\%, and is similar for positive and negative particles.
The most important correction when calculating the particle ratios is that related to the track finding inefficiency,  as different interaction cross-sections and decays in flight mean that this effect does not in general cancel.  All correction factors are taken from simulation, and are applied bin-by-bin, after which the particle ratios are determined. The corrections typically lead to a change of  less than a relative 10\% on the ratios.

The analysis procedure is validated on simulated events in which the measured ratios are compared with those expected from generator level. A $\chi^2$ is formed over all the $\eta$ bins at low \pt, summed over the different-particle ratios.  Good agreement is found for the same-particle ratios over all $\eta$ and \pt, and for the different-particle ratios at mid and high \pt.
Discrepancies are however observed at low \pt for the different-particle ratios,  which are attributed to 
imperfections in the PID reweighting procedure for this region. The $\chi^2$ in the low \pt bin is then minimised by applying charge-independent scaling factors of  $1.33$ ($1.10$) and $0.90$ ($0.86$) for the proton and kaon efficiencies, respectively, at 0.9~TeV (7~TeV). 
An uncertainty of $\pm 0.11$ is assigned to the scaling factors, uncorrelated bin-to-bin,  in order to obtain $\chi^2/{\rm ndf} \approx 1$ at both energy points.  
This uncertainty is fully correlated between positive and negative tracks.
Although no bias is observed at mid and high \pt, an additional relative uncertainty of $\pm 0.03$ is assigned to the proton and kaon efficiencies for these bins to yield an acceptable scatter (\ie\ $\chi^2/{\rm ndf} \approx 1$). This uncertainty is also taken to be uncorrelated bin-to-bin, but fully correlated between positive and negative tracks.  
The scaling factors and uncertainties from these studies
are adopted for the analysis of the data.

\section{Systematic uncertainties}
\label{sec:systs}

The contribution to the systematic uncertainty of all effects considered is summarised in Tables~\ref{tab:syst_like_09} and~\ref{tab:syst_like_7} for the same-particle ratios, and in  Tables~\ref{tab:syst_unlike_09} and~\ref{tab:syst_unlike_7} for the different-particle ratios.  

The dominant uncertainty is associated with the understanding of the PID performance.  Each element in the identification matrix (Eq.~\ref{eq:id_matrix}), is smeared by a Gaussian of width corresponding to the uncertainty in the identification (or misidentification) efficiency of that element, and the full set of particle ratios is recalculated.  This uncertainty is the sum in quadrature of the statistical error from the calibration sample after reweighting, as discussed in Sect.~\ref{sec:pid}, and the additional uncertainty assigned after the analysis validation, described in Sect.~\ref{sec:analysis}.
 The procedure is repeated many times and the width of the resulting distributions is assigned as the systematic uncertainty.  
As can be seen in Tables~\ref{tab:syst_like_09}--\ref{tab:syst_unlike_7} there is a large range in the magnitude of this contribution.
The uncertainty is smallest at high \pt and $\eta$, on account of the distribution of the events in the calibration sample.
For each observable the largest value  is found in the lowest $\eta$ bin at mid-\pt.  If this bin and the lowest $\eta$ bin at low $\pt$ are discounted, the variation in uncertainty of the remainder of the acceptance is much smaller, being typically a factor of two or three.

Knowledge of the interaction cross-sections and the amount of  material encountered by particles in traversing the spectrometer is necessary
to determine the fraction of particles that cannot be reconstructed due to having undergone a strong interaction. 
The interaction cross-sections as implemented in the LHCb simulation agree with measurements~\cite{COMPAS} over the momentum range of interest to a precision of around 20\% for protons and kaons, and 10\% for pions.  The material description up to and including the tracking detectors is correct within a tolerance of 10\%.   The effect of these uncertainties is propagated through in the calculation of the track loss for each particle type from strong interaction effects.

The detection efficiency of positive and negative tracks need not be identical due to the fact that each category
 is swept by the dipole field, on average, to  different regions of the spectrometer.    Studies using $J/\psi \to \mu^+\mu^-$ decays in which one track is selected by muon chamber information alone constrain any charge asymmetry in the track reconstruction efficiency to be less than 1.0 (0.5)\% for the 0.9 (7)~TeV  data. These 
values are used to assign systematic uncertainties on the particle ratios. The identification efficiencies in the RICH system are measured separately for each charge, and so this effect is accounted for in the inputs to the analysis.  A cross-check that there are no significant reconstruction asymmetries left unaccounted for is provided by a comparison of  the results obtained with the two polarity settings of the dipole magnet.  Consistent results are found for all observables.

A possible source of bias arises from the contribution of `ghost' tracks; these are tracks which have no correspondence with the trajectory of any charged particle in the event, but are reconstructed from the incorrect association of hit points in the tracking detectors.
Systematic uncertainties are therefore assigned in each $(p_T, \eta)$ bin for each category of ratio by subtracting the estimated contribution of ghost tracks for each particle assignment, and determining the resulting shifts in the calculated ratios.  A sample enriched in ghost tracks can be obtained by selecting tracks where the number of hits associated with the track in the TT detector is significantly less than that expected for a particle with that trajectory.   Comparison of the fraction of tracks of this nature in data and simulation is used to determine the ghost-track rate in data by scaling the known rate in simulation.   This exercise is performed independently for identified tracks which are above and below the Cherenkov threshold in the RICH system. The contamination from ghost tracks is lower in the above-threshold category since the presence of photodetector hits is indicative of a genuine track.    The total ghost-track fraction for pions and kaons  is found to be typically below 1\%, rising to around 2\% in certain bins.   The ghost-track fraction for protons rises to 5\% in some bins, on account of the larger fraction of this particle type lying below the  RICH threshold.
The charge asymmetry for this background is found to be small and the assigned systematic uncertainty is in general around 0.1\%.
To provide further confirmation that ghost tracks are not a significant source of bias  the analysis is repeated with different cut values on the track-fit $\chi^2/{\rm ndf}$ and stable results are found.

Clones are suppressed by the requirement that only one track is retained from pairs of tracks that have very similar momentum.  The analysis is repeated with the requirement removed, and negligible changes are seen for all observables. 

Contamination from non-prompt particles induces a small uncertainty in the measurement, as this source of background is at a low level  and cancels to first order in the ratios. The error is assigned by repeating the analysis and doubling the assumed charge asymmetry of these tracks compared with the value found from the simulation.  No significant variations are observed when the analysis is repeated with different cut values on the prompt-track selection variable $\chi^2_{\rm IP}$.   

The total systematic uncertainty for each observable is obtained by summing in quadrature the individual contributions in each (\pt, $\eta$) bin.   In general, the systematic uncertainty  is significantly  larger than the statistical uncertainty, with the largest contribution coming from the knowledge of the PID performance, which is limited by the size of the calibration sample.

\begin{table}
\begin{center}
\caption{\small Range of systematic uncertainties, in percent, for same-particle ratios at $\sqrt{s}=0.9$ TeV.}\label{tab:syst_like_09}
\begin{tabular}{l|r|r|r}
 & \multicolumn{1}{c|}{\hspace*{0.1cm}$\bar{p}/{p}$} & \multicolumn{1}{c|}{\hspace*{0.3cm}$K^-/K^+$} &\multicolumn{1}{c}{\hspace*{0.3cm}$\pi^-/\pi^+$} \\ \hline
PID                           & 7.5 $-$ 46.7                                       & 4.9 $-$ 42.4                                 & 0.8 $-$ 6.0 \\ 
Cross-sections         & 0.2 $-$ \hspace*{0.2cm}1.6             & 0.1 $-$ \hspace*{0.2cm}1.5        &  $<$0.1 $-$ 0.8 \\ 
Detector material     & 0.1 $-$ \hspace*{0.2cm}0.8             & 0.1 $-$ \hspace*{0.2cm}0.7        & $<$0.1 $-$ 0.8 \\  
Ghosts                      & $<$0.1 $-$ \hspace*{0.2cm}0.1      &$< $0.1 $-$ \hspace*{0.2cm}0.1  & $<$0.1 $-$ 0.1 \\
Tracking asymmetry & 1.0                                                     &1.0                                                 & 1.0 \\
Non-prompt              & $<$0.1 $-$ \hspace*{0.2cm}0.2      &$<$0.1 $-$ \hspace*{0.2cm}0.1   & $<$0.1 $-$ 0.1 \\ \hline
Total                         & 7.7 $-$ 46.7                                       & 5.0 $-$ 42.4                                 &  1.3 $-$ 6.0 \\ 
\end{tabular}
\end{center}
\end{table}

\begin{table}
\begin{center}
\caption{\small Range of systematic uncertainties, in percent, for same-particle ratios at $\sqrt{s}=7$ TeV.}\label{tab:syst_like_7}
\begin{tabular}{l|r|r|r}
 & \multicolumn{1}{c|}{\hspace*{0.1cm}$\bar{p}/{p}$} & \multicolumn{1}{c|}{\hspace*{0.3cm}$K^-/K^+$} &\multicolumn{1}{c}{\hspace*{0.3cm}$\pi^-/\pi^+$} \\ \hline
PID                           & 3.4 $-$ 26.4                                       & 2.0 $-$ 15.8                                 & 0.6 $-$ 2.7 \\ 
Cross-sections         & 0.3 $-$ \hspace*{0.2cm}1.8             & 0.3 $-$ \hspace*{0.2cm}0.7        &  $<$0.1 $-$ 0.2 \\ 
Detector material     & 0.2 $-$ \hspace*{0.2cm}0.9             & 0.1 $-$ \hspace*{0.2cm}0.4        & $<$0.1 $-$ 0.2 \\  
Ghosts                      & $<$0.1 $-$ \hspace*{0.2cm}0.4      &$< $0.1 $-$ \hspace*{0.2cm}0.1  & $<$0.1  \\
Tracking asymmetry & 0.5                                                     &0.5                                                 & 0.5 \\
Non-prompt              & $<$0.1 $-$ \hspace*{0.2cm}0.2      &$<$0.1 $-$ \hspace*{0.2cm}0.1   & $<$0.1 $-$ 0.1 \\ \hline
Total                         & 3.5 $-$ 26.5                                       & 2.1 $-$ 15.8                                 &  0.8 $-$ 2.8 \\ 
\end{tabular}
\end{center}
\end{table}

\begin{table}
\begin{center}
\caption{\small Range of systematic uncertainties, in percent, for different-particle ratios at $\sqrt{s}=0.9$ TeV.}\label{tab:syst_unlike_09}
\begin{tabular}{l|r|r|r}
 &  $(p + \bar{p})/(\pi^+ + \pi^-)$  & $(K^+ + K^-)/(\pi^+ + \pi^-)$ &  $(p + \bar{p})/(K^+ + K^-)$ \\ \hline
PID                           &  10.2 $-$ 63.7       \hspace*{0.5cm}                            & 8.1 $-$ 46.8                              \hspace*{0.8cm} & 5.9 $-$ 42.6  \hspace*{0.6cm} \\ 
Cross-sections         & 0.1 $-$ \hspace*{0.2cm}1.6         \hspace*{0.5cm}      & 0.4 $-$ \hspace*{0.2cm}1.3      \hspace*{0.8cm}  & 0.2 $-$  \hspace*{0.2cm}2.4 \hspace*{0.6cm}\\ 
Detector material     & $<$0.1 $-$ \hspace*{0.2cm}0.8  \hspace*{0.5cm}      & 0.2 $-$ \hspace*{0.2cm}0.7      \hspace*{0.8cm}  & 0.1 $-$  \hspace*{0.2cm}1.2 \hspace*{0.6cm} \\  
Ghosts                      & $<$0.1 $-$ \hspace*{0.2cm}0.1 \hspace*{0.5cm}      &$< $0.1 $-$ \hspace*{0.2cm}0.1  \hspace*{0.8cm}& $<$0.1 $-$  \hspace*{0.2cm}0.1  \hspace*{0.6cm} \\
Tracking asymmetry & $<$0.1 \hspace*{0.5cm}& $<$0.1 \hspace*{0.8cm}& $<$0.1 \hspace*{0.6cm} \\
Non-prompt              & $<$0.1 $-$ \hspace*{0.2cm}0.2  \hspace*{0.5cm}     &              0.1                                  \hspace*{0.8cm}  & $<$0.1 $-$  \hspace*{0.2cm}0.1 \hspace*{0.6cm} \\ \hline
Total                         & 10.2 $-$ 63.7                                  \hspace*{0.5cm}    & 8.6 $-$ 46.8                                \hspace*{0.8cm}  &  6.0 $-$ 42.6 \hspace*{0.6cm} \\ 
\end{tabular}
\end{center}
\end{table}

\begin{table}
\begin{center}
\caption{\small Range of systematic uncertainties, in percent, for different-particle ratios at $\sqrt{s}=7$ TeV.}\label{tab:syst_unlike_7}
\begin{tabular}{l|r|r|r}
 &  $(p + \bar{p})/(\pi^+ + \pi^-)$  & $(K^+ + K^-)/(\pi^+ + \pi^-)$ &  $(p + \bar{p})/(K^+ + K^-)$ \\ \hline
PID                           & 5.9 $-$ 31.1                                \hspace*{0.5cm}      & 4.6 $-$ 26.6                                  \hspace*{0.8cm} & 3.7 $-$ 16.1 \hspace*{0.6cm} \\ 
Cross-sections         & 0.3 $-$ \hspace*{0.2cm}2.2         \hspace*{0.5cm}      & 1.2 $-$ \hspace*{0.2cm}1.5       \hspace*{0.8cm}  & 0.2 $-$  \hspace*{0.2cm}2.1 \hspace*{0.6cm}\\ 
Detector material     & 0.2 $-$ \hspace*{0.2cm}1.1       \hspace*{0.5cm}      & 0.6 $-$ \hspace*{0.2cm}0.8        \hspace*{0.8cm}  & 0.1 $-$  \hspace*{0.2cm}1.0 \hspace*{0.6cm} \\  
Ghosts                      & $<$0.1 $-$ \hspace*{0.2cm}0.3 \hspace*{0.5cm}      &$< $0.1 $-$ \hspace*{0.2cm}0.3  \hspace*{0.8cm}& $<$0.1 $-$  \hspace*{0.2cm}0.2  \hspace*{0.6cm} \\
Tracking asymmetry & $<$0.1 \hspace*{0.5cm}& $<$0.1 \hspace*{0.8cm}& $<$0.1 \hspace*{0.6cm} \\
Non-prompt              & $<$0.1 $-$ \hspace*{0.2cm}0.3  \hspace*{0.5cm}     &  0.1 $-$ \hspace*{0.2cm}0.2       \hspace*{0.8cm}  & $<$0.1 $-$  \hspace*{0.2cm}0.2 \hspace*{0.6cm} \\ \hline
Total                         & 6.0 $-$ 31.1                                  \hspace*{0.5cm}    & 4.8 $-$ 26.7                                   \hspace*{0.8cm}  &  3.7 $-$ 16.2 \hspace*{0.6cm} \\ 
\end{tabular}
\end{center}
\end{table}

\section{Results}
\label{sec:results}

\begin{figure}
 \begin{center}
 \subfigure{\includegraphics[width=0.48\textwidth]{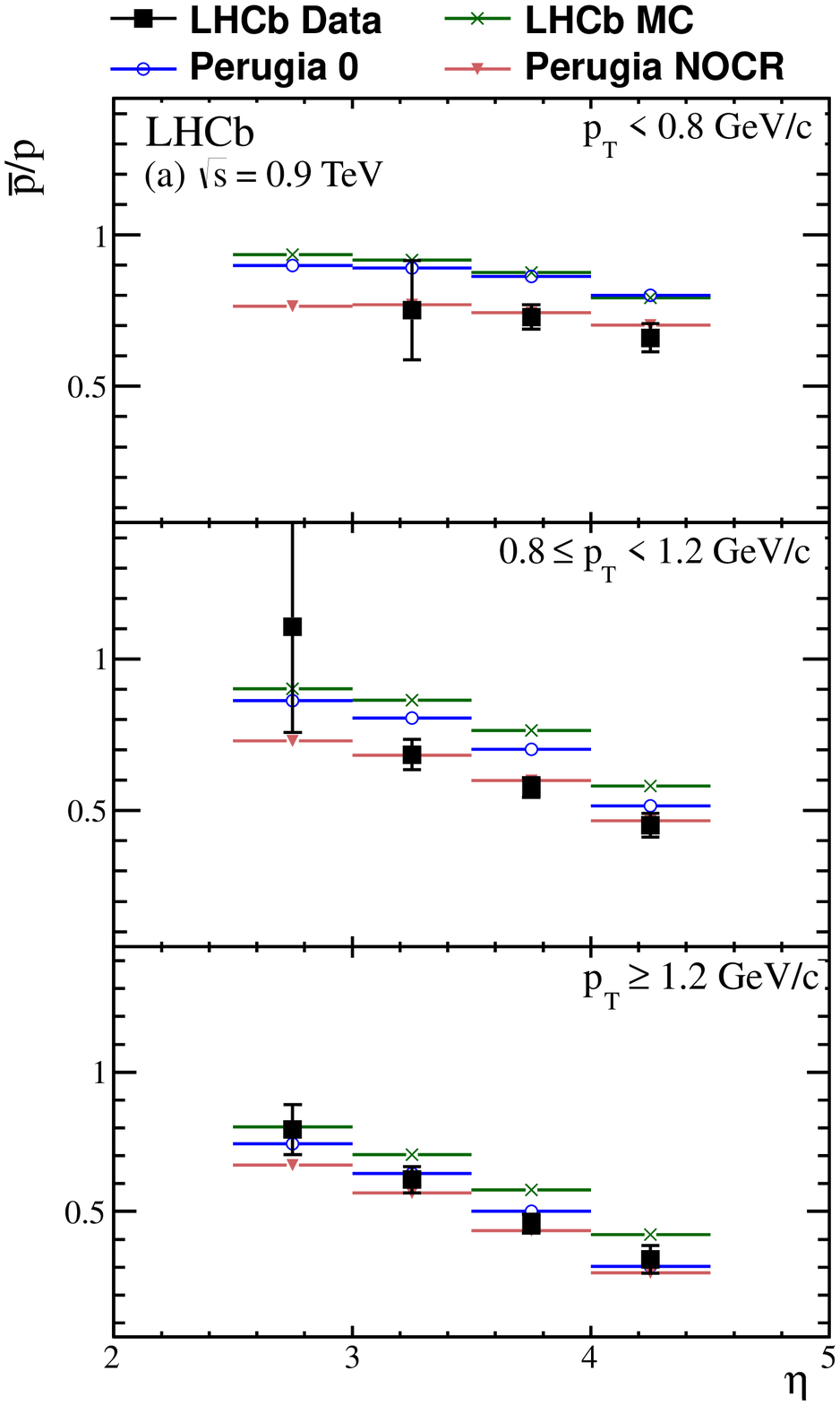}}
 \subfigure{\includegraphics[width=0.48\textwidth]{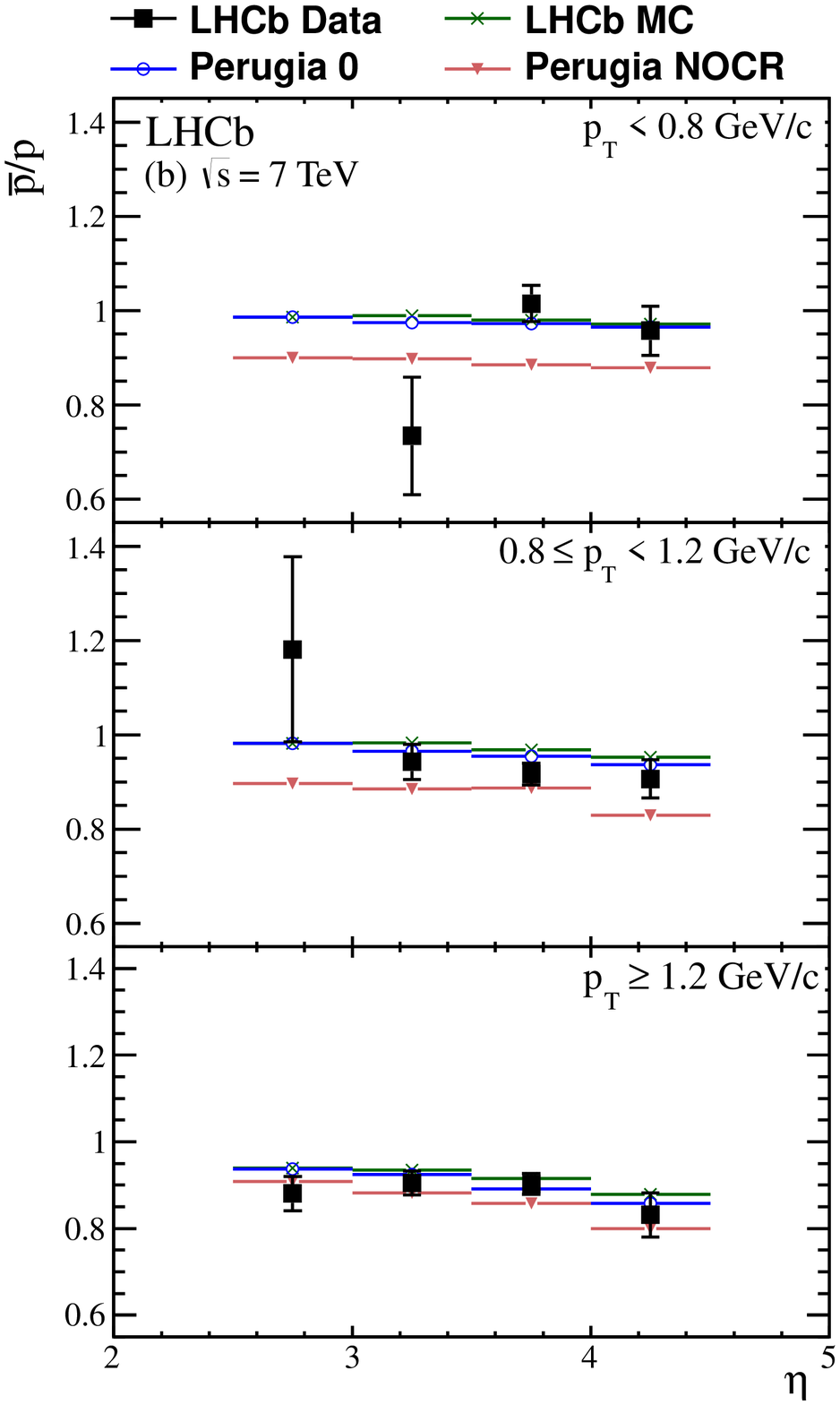}}
  \caption{\small Results for the $\bar{p}/p$ ratio at 0.9~TeV (a) and 7~TeV (b). }
 \label{fig:proton_eta_final}
 \end{center}
\end{figure}

\begin{figure}
 \begin{center}
 \subfigure{\includegraphics[width=0.48\textwidth]{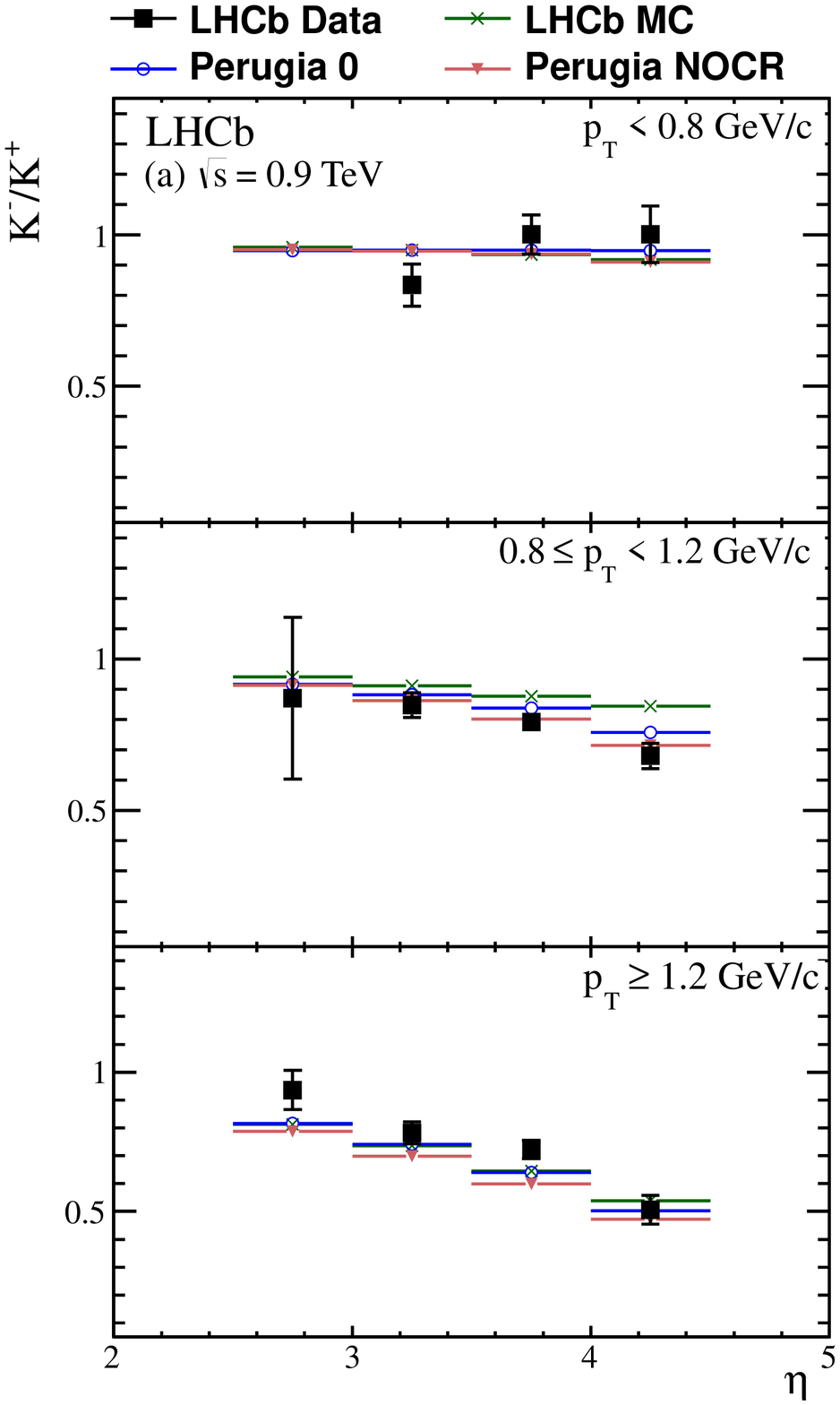}}
 \subfigure{\includegraphics[width=0.48\textwidth]{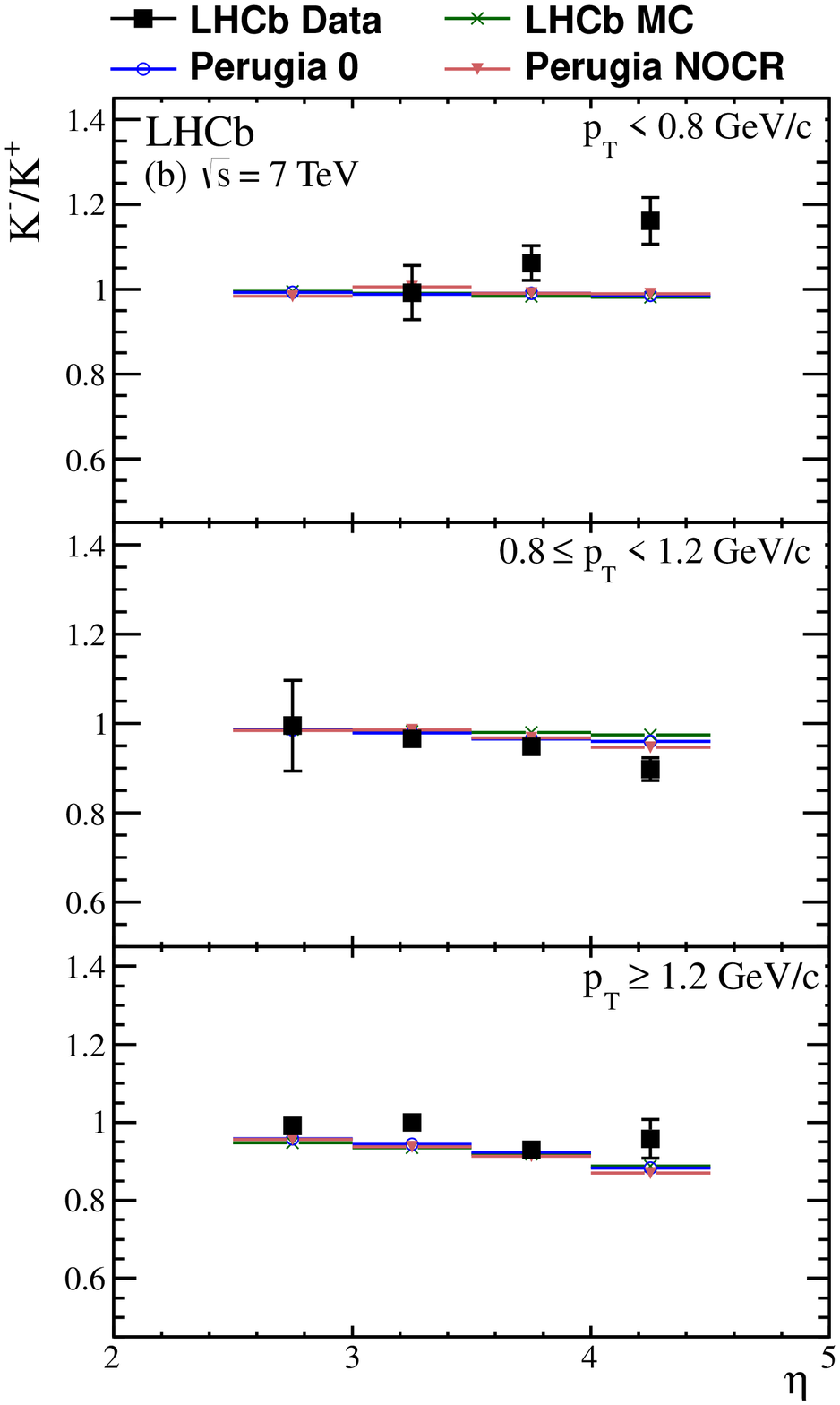}}
  \caption{\small Results for the $K^{-}/K^{+}$  ratio at 0.9~TeV (a) and 7~TeV (b).}
 \label{fig:kaon_eta_final}
 \end{center}
\end{figure}

\begin{figure}
 \begin{center}
 \subfigure{\includegraphics[width=0.48\textwidth]{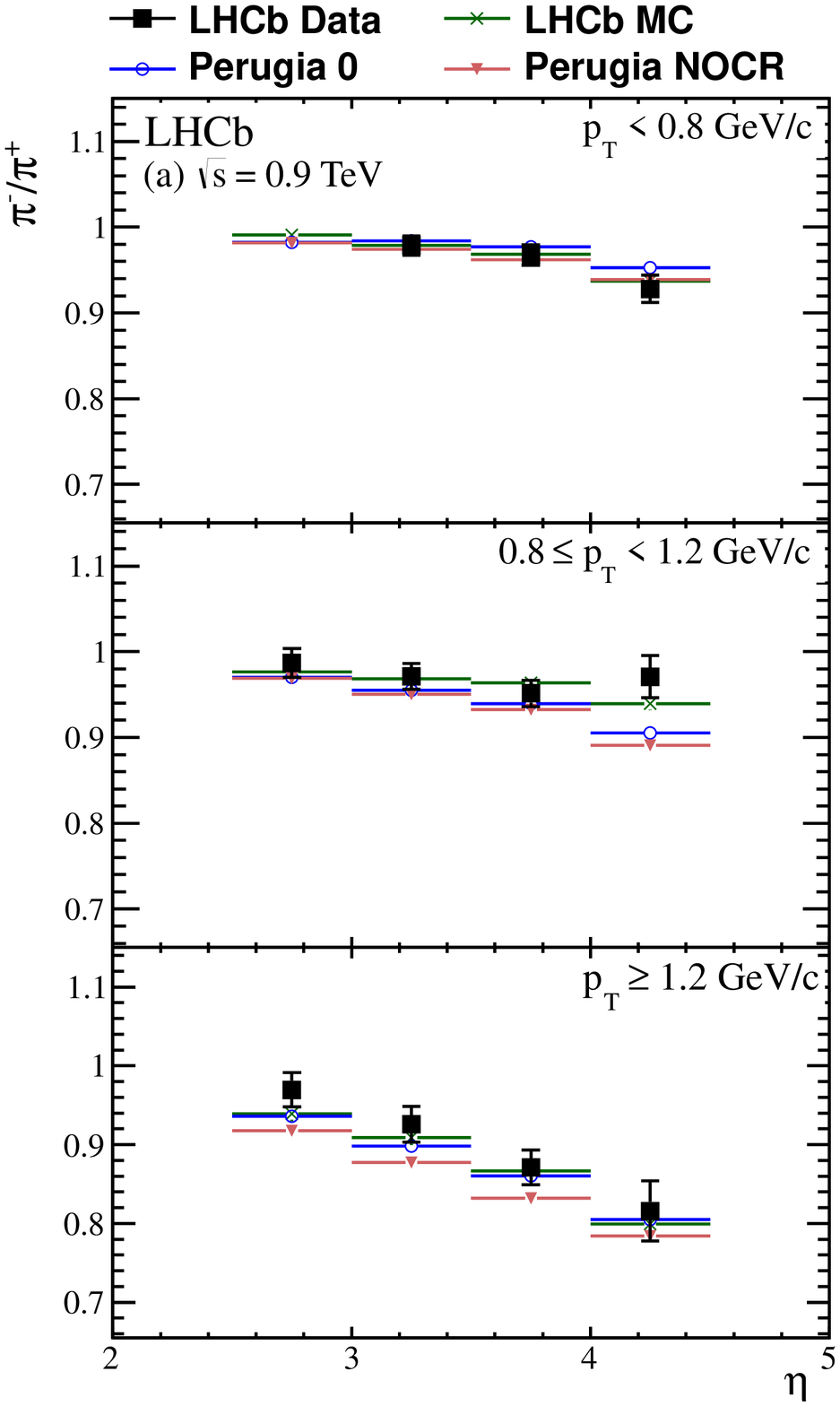}}
 \subfigure{\includegraphics[width=0.48\textwidth]{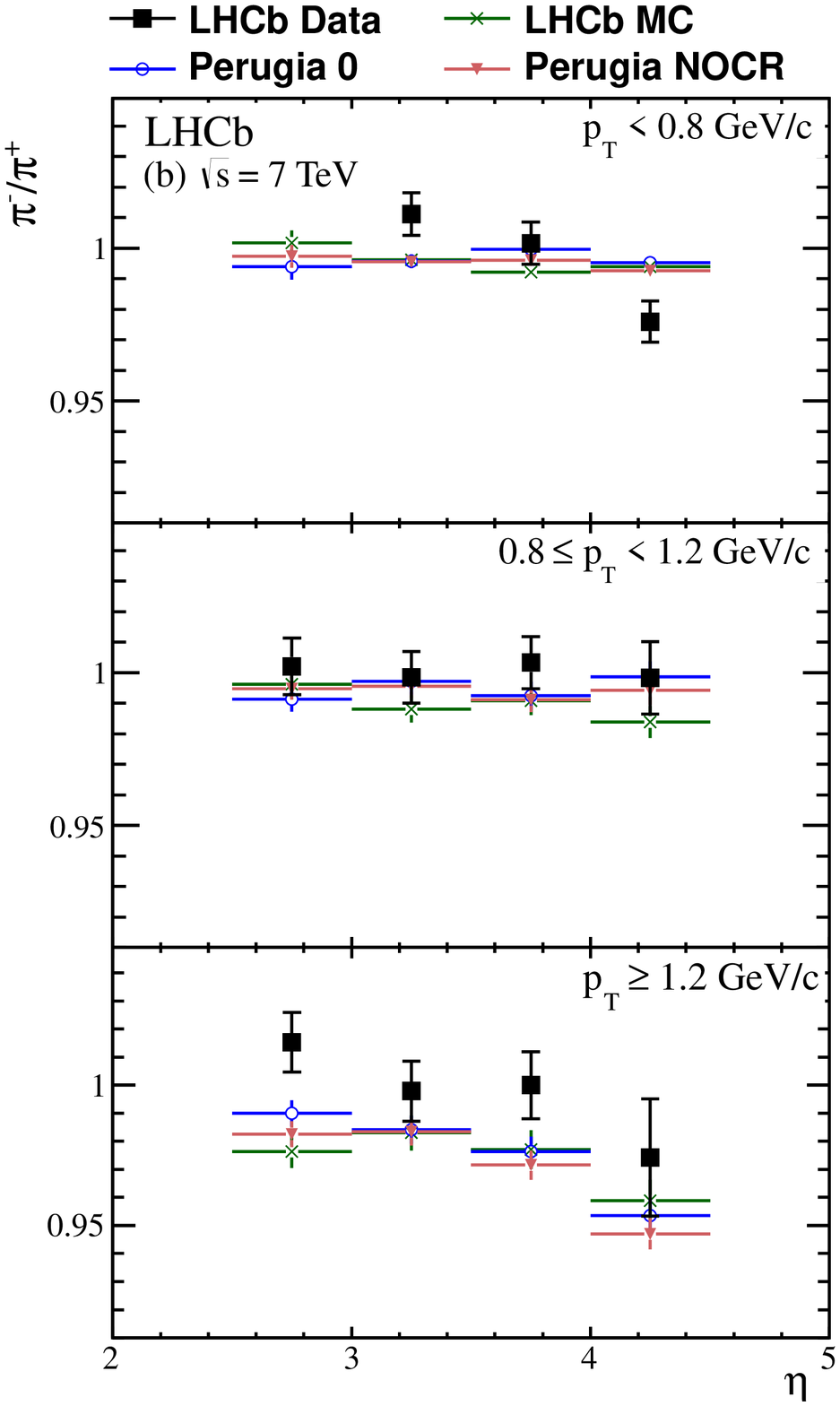}}
  \caption{\small Results for the $\pi^{-}/\pi^{+}$ ratio at 0.9~TeV (a) and 7~TeV (b). }
 \label{fig:pion_eta_final}
 \end{center}
\end{figure}

\begin{figure}[htbp]
 \begin{center}
  \subfigure{\includegraphics[width=0.48\textwidth]{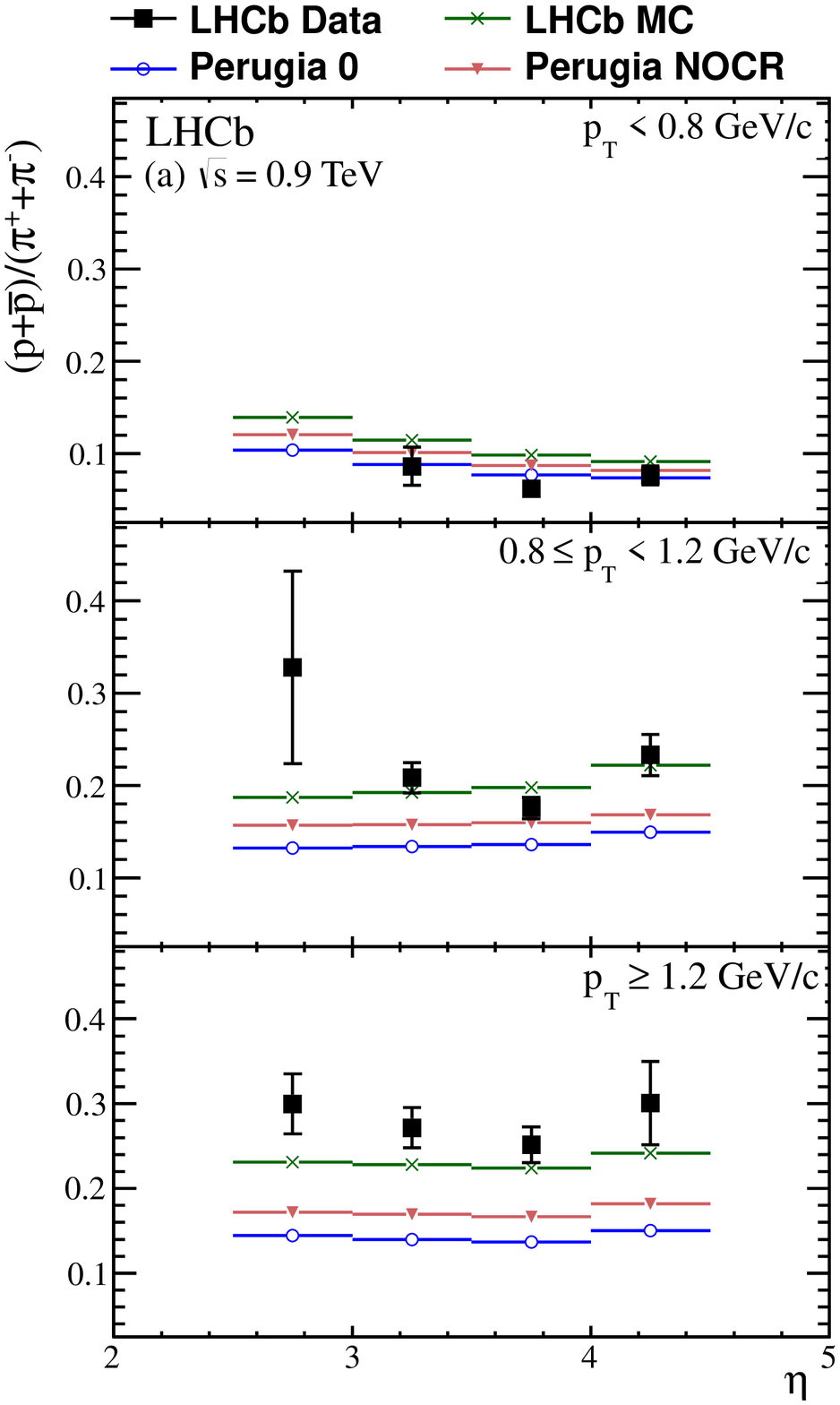}}
 \subfigure{\includegraphics[width=0.48\textwidth]{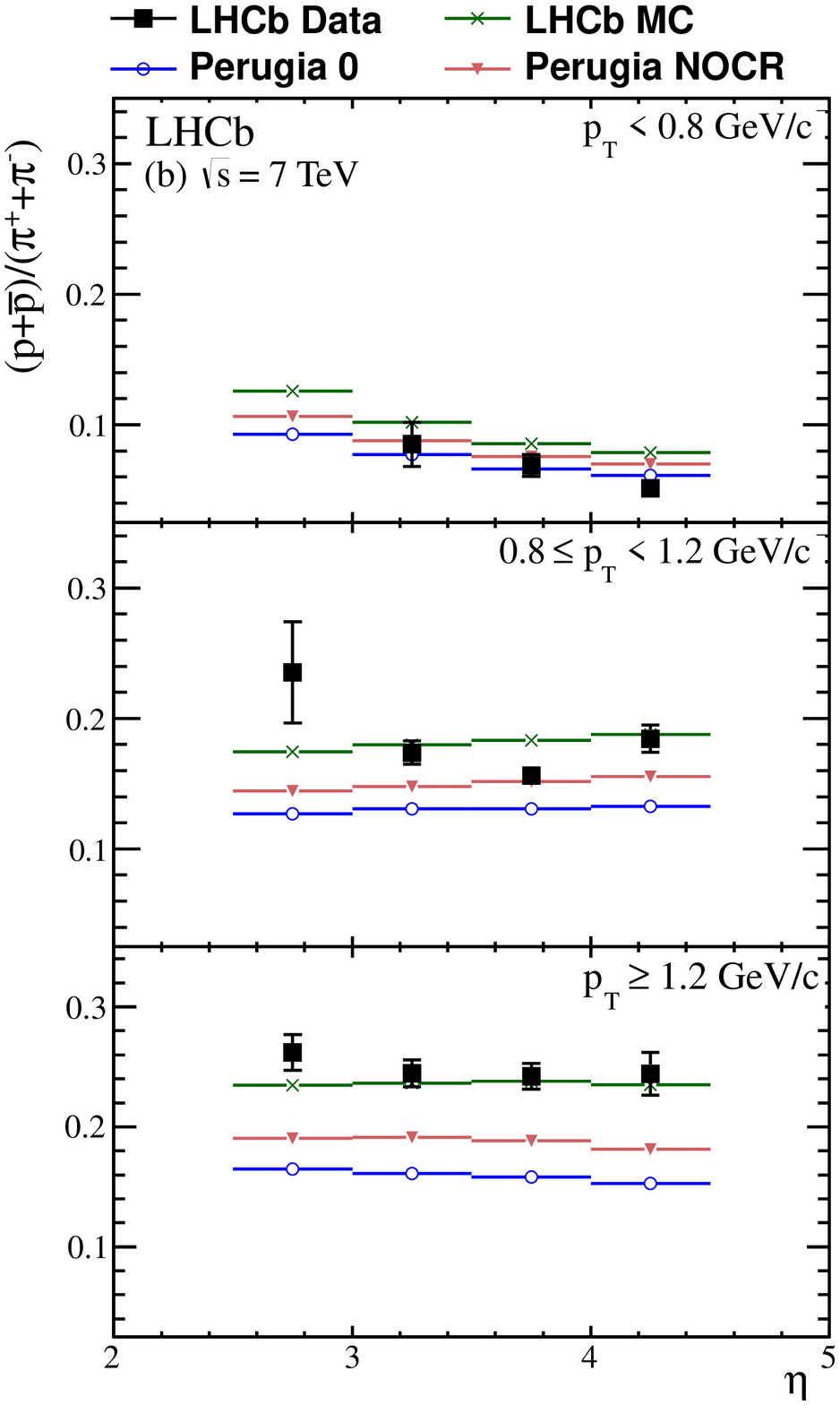}}
  \caption{\small Results for the $(p + \bar{p})/(\pi^+ + \pi^-)$ ratio at 0.9~TeV (a) and 7~TeV (b). }
 \label{fig:ppi_eta_final}
 \end{center}
\end{figure}

The measurements of the same-particle ratios are plotted in Figs.~\ref{fig:proton_eta_final}, ~\ref{fig:kaon_eta_final} and~\ref{fig:pion_eta_final}, and those of the different-particle ratios in Figs.~\ref{fig:ppi_eta_final},~\ref{fig:Kpi_eta_final} and~\ref{fig:pK_eta_final}.  The numerical values can be found in Appendix~\ref{sec:tabresults}.  Also shown are the predictions of several \textsc{Pythia6.4} generator settings, or `tunes': LHCb~MC\cite{VANYA}, Perugia~0 and Perugia NOCR\cite{PERUGIA}.  At 0.9~TeV the $\bar{p}/p$ ratio falls from around 0.8 at low $\eta$ to  around 0.4 in the highest \pt and $\eta$ bin.  At this energy point there is a significant spread between models for the Monte Carlo predictions, with the data lying significantly below the LHCb~MC and Perugia~0 expectations, but close to those of Perugia NOCR.  At higher energy the $\bar{p}/p$ ratio is higher and varies more slowly, in good agreement with LHCb~MC and Perugia~0 and less so with Perugia~ NOCR.  The $K^-/K^+$ and $\pi^-/\pi^+$ ratios also differ from unity, most noticeably at high \pt  and high $\eta$.  This behaviour is in general well modelled by all the generator tunes, which give similar predictions for these observables.  
Small discrepancies are observed at 7~TeV for $K^-/K^+$  at low \pt, and $\pi^-/\pi^+$ at high \pt.
When comparing the measurements and predictions for the different-particle ratios the most striking differences occur for $(p + \bar{p})/(\pi^+ + \pi^-)$ and $(K^+ + K^-)/(\pi^+ + \pi^-)$, where there is a tendency for the data to lie significantly higher than the Perugia~0 and NOCR expectations.  The agreement with the LHCb~MC for these observables is generally good.

\begin{figure}[htbp]
 \centering
 \includegraphics[width=12cm]{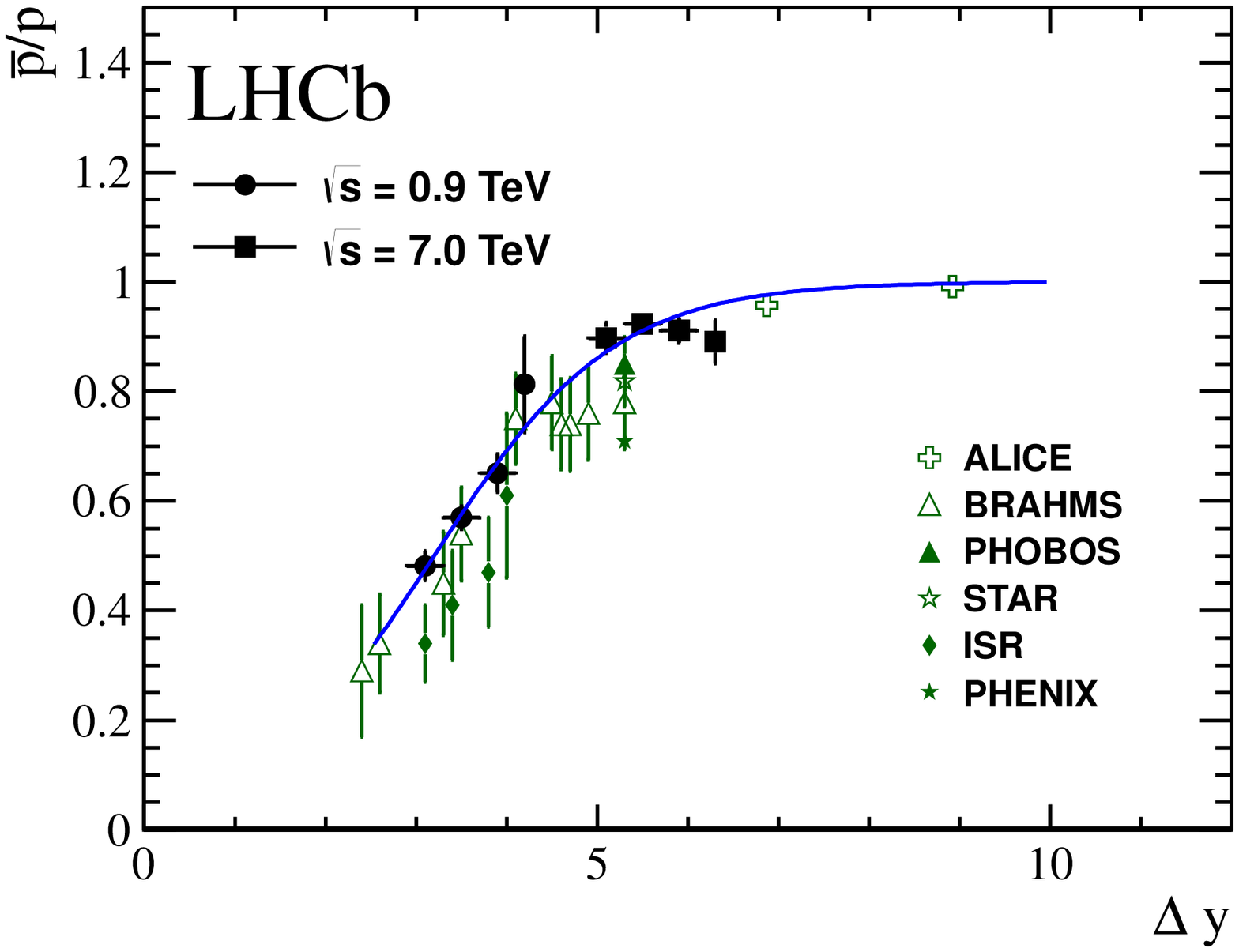}
  \caption{\small Results for the $\bar{p}/p$ ratio against the rapidity loss $\Delta y$ from LHCb. Results from other experiments are also shown~\cite{ALICEPBARP,ISR,BRAHMS, PHENIX,PHOBOS,STAR}.  Superimposed is a fit to the LHCb and ALICE~\cite{ALICEPBARP} measurements that is described in the text.}
 \label{fig:proton_DeltaY}
\end{figure}

It is instructive to consider the $\bar{p}/p$ results as a function of rapidity loss, $\Delta y \equiv y_{\rm beam} - y$, where $y_{\rm beam}$ is the rapidity of the protons in the LHC beam which travels forward in the spectrometer ($y_{\rm beam} = 6.87$ at 0.9~TeV and 8.92 at 7~TeV).  For the same-particle ratios it is possible to determine the rapidity value to which the measurement in each $\eta$ bin corresponds.  In each bin the mean and RMS spread of the rapidity of the tracks in the analysis sample is determined. 
Correlations are accounted for, but these are in general negligible as the uncertainties are dominated by the PID errors, which for these observables are statistical in nature.
A small correction is applied to this mean, obtained from Monte Carlo, to account for the distortion to the unbiased spectrum that is induced by the reconstruction and PID requirements. The values of the mean and RMS spread of the rapidities for $\bar{p}/p$ can be found in Appendix~\ref{sec:tabresults},
together with those of $K^-/K^+$ and $\pi^-/\pi^+$.  
As no evidence is seen of any \pt dependence in the distribution of the $\bar{p}/p$ results against $\Delta y$ the measurements in each $\eta$ bin at each energy point are integrated over \pt, with the uncertainties on the individual values of the ratios used to determine the weights of each input entering into the mean. 
The mean $\bar{p}/p$ ratios are given as a function of $\Delta y$ in Table~\ref{tab:delta_y_ptavg_proton} and plotted in Fig.~\ref{fig:proton_DeltaY}, with the results from other experiments~\cite{ALICEPBARP,ISR,BRAHMS,PHENIX,PHOBOS,STAR}  superimposed. 
The LHCb results cover a wider range of $\Delta y$ than any other single experiment and significantly improve the precision of the measurements in the region $\Delta y < 6.5$.

\begin{table}
\begin{center}
\caption{\small Results for $\bar{p}/p$ ratio  integrated over \pt in $\eta$ bins as a function of the rapidity loss $\Delta y$. } \label{tab:delta_y_ptavg_proton}
\begin{tabular}{l|c|c|c}
$\sqrt{\mathrm{s}}$ & $\eta$ range & $\Delta y$ & Ratio\\
\hline
$0.9\,\mathrm{TeV}$
& $4.0-4.5$ & $ 3.1 \pm 0.2$ & $ 0.48 \pm 0.03$\\
& $3.5-4.0$ & $ 3.5 \pm 0.2$ & $ 0.57 \pm 0.02$\\
& $3.0-3.5$ & $ 3.9 \pm 0.2$ & $ 0.65 \pm 0.03$\\
& $2.5-3.0$ & $ 4.3 \pm 0.1$ & $ 0.81 \pm 0.09$\\
\hline
$7\,\mathrm{TeV}$
& $4.0-4.5$ & $ 5.1 \pm 0.2$ & $ 0.90 \pm 0.03$\\
& $3.5-4.0$ & $ 5.5 \pm 0.2$ & $ 0.92 \pm 0.02$\\
& $3.0-3.5$ & $ 5.9 \pm 0.2$ & $ 0.91 \pm 0.02$\\
& $2.5-3.0$ & $ 6.3 \pm 0.1$ & $ 0.89 \pm 0.04$\\
\end{tabular}
\end{center}
\end{table}

Within the Regge model, baryon production at high energy is driven by Pomeron exchange and baryon transport by string-junction exchange~\cite{THEORY5}. Assuming this picture the $\Delta y$ dependence of the $\bar{p}/p$ ratio approximately follows the form $1/ \left( 1 + C \exp [(\alpha_J - \alpha_P) \Delta y] \right)$, where $C$ determines the relative contributions of the two mechanisms, and $\alpha_J$ ($\alpha_P$) is the intercept of the string junction (Pomeron) Regge trajectory. Figure~\ref{fig:proton_DeltaY} shows the results of fitting this expression to both the LHCb and, in order to constrain the high $\Delta y$ region,  the ALICE data.  Both $C$ and $(\alpha_J - \alpha_P)$ are free parameters of the fit and are determined to be $22.5 \pm 6.0$ and $-0.98 \pm 0.07$ respectively with a $\chi^2$/ndf of $8.7/8$.  Taking $\alpha_P = 1.2$ ~\cite{KAIDALOV} suggests a low value of $\alpha_J$, significantly below the $\alpha_J \approx 0.5$ expected if the string-junction intercept is associated with that of the standard Reggeon (or meson). The value of $\alpha_J \approx 0.9$ which would be expected if the string junction is associated with the Odderon\cite{THEORY8} is excluded using this fit model.
The same conclusion applies if the LHCb and ALICE $\bar{p}/p$ ratio values are fitted with an alternative parameterisation~\cite{THEORY6b}  $C^\prime \cdot (s [{\rm GeV}^2])^{(\alpha_J - \alpha_P)/2}\cdot \cosh [y (\alpha_J - \alpha_P) ]$, which yields the results $C^\prime = 10.2 \pm 1.8$, 
$(\alpha_J - \alpha_P) = -0.86 \pm 0.05$ with a $\chi^2$/ndf of $10.2/8$.


\begin{figure}[htbp]
 \begin{center}
  \subfigure{\includegraphics[width=0.48\textwidth]{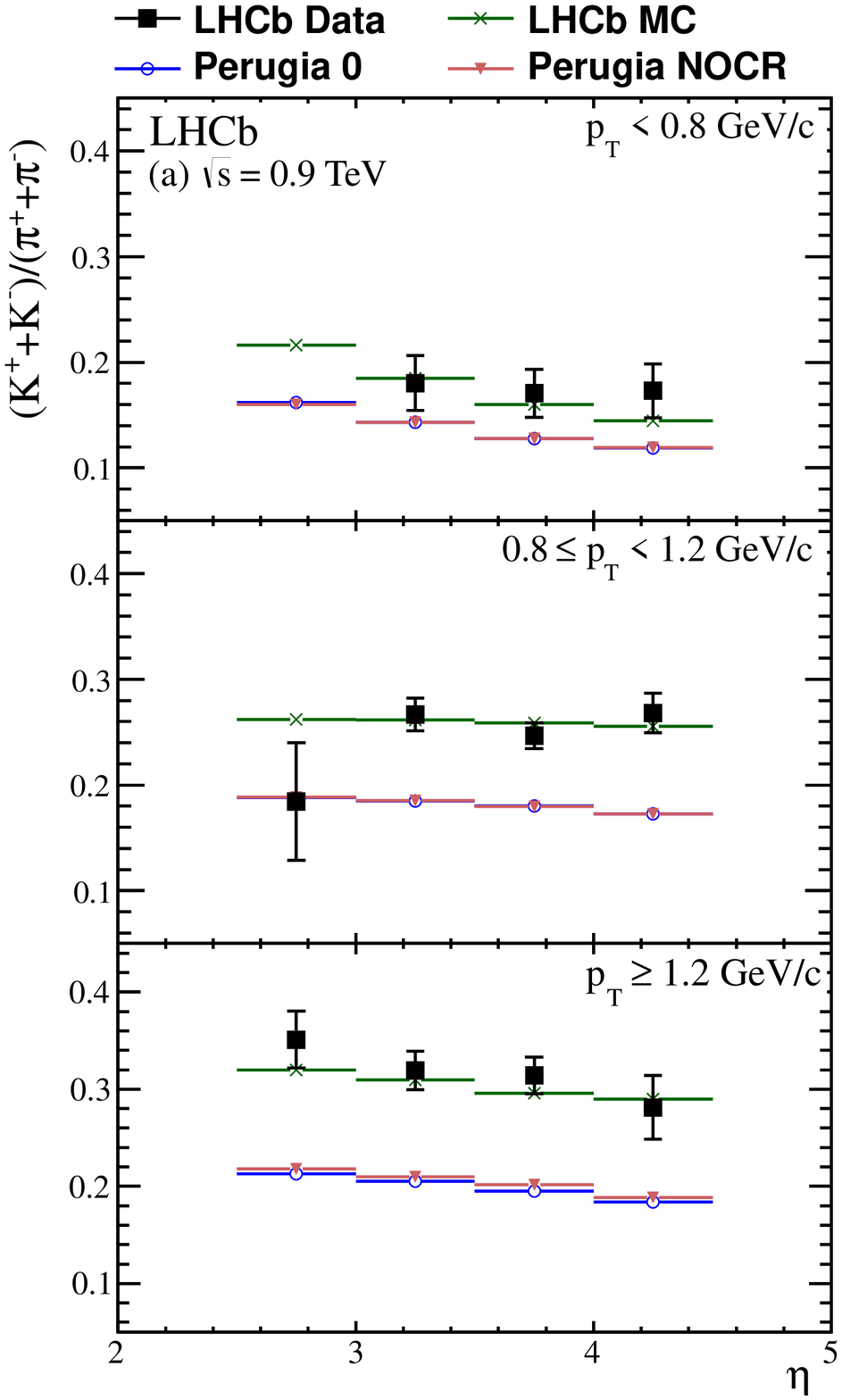}}
 \subfigure{\includegraphics[width=0.48\textwidth]{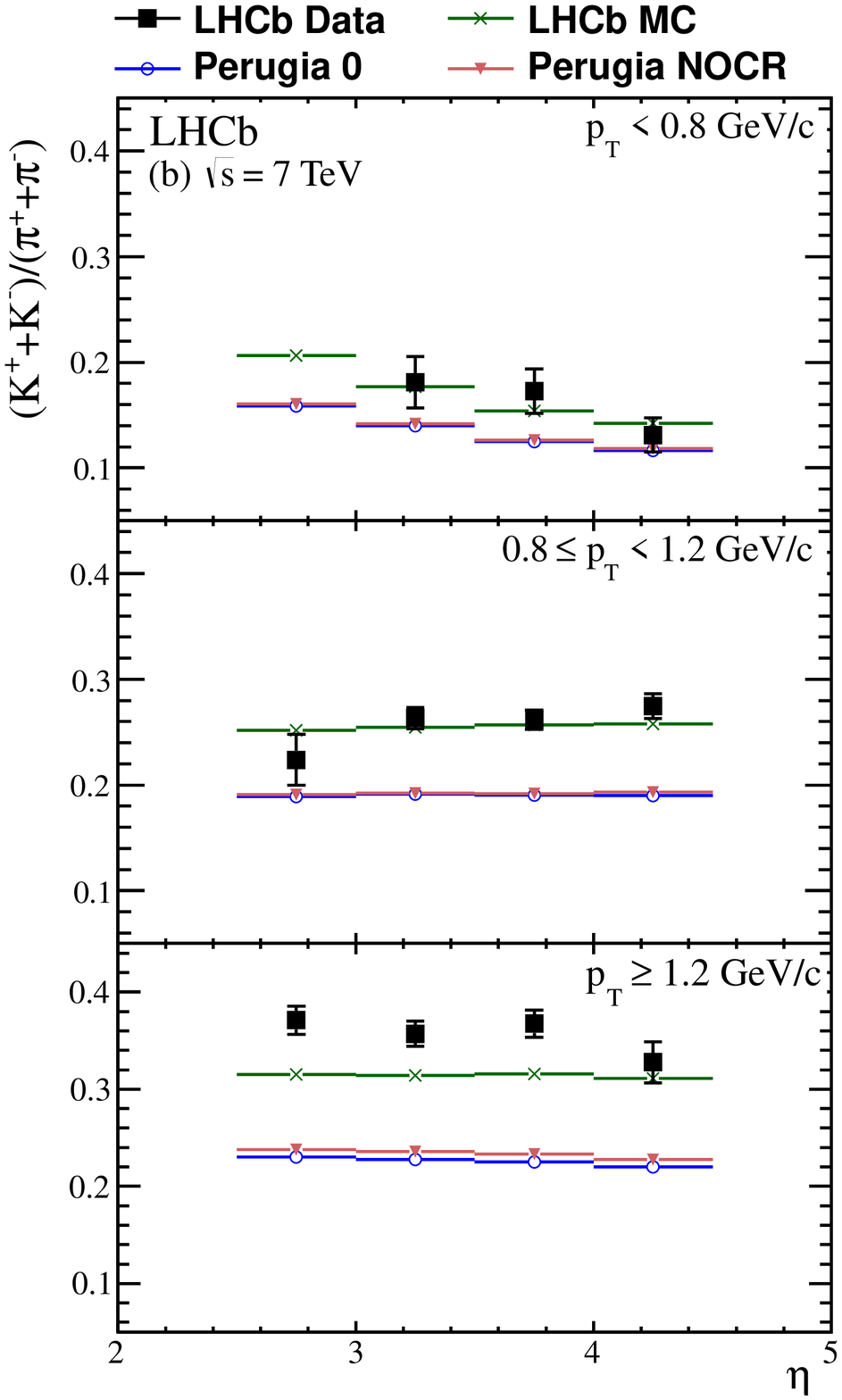}}
  \caption{\small Results for the $(K^+ + K^-)/(\pi^+ + \pi^-)$ ratio at 0.9~TeV (a) and 7~TeV (b).}
 \label{fig:Kpi_eta_final}
 \end{center}
\end{figure}

\begin{figure}[htbp]
 \begin{center}
 \subfigure{\includegraphics[width=0.48\textwidth]{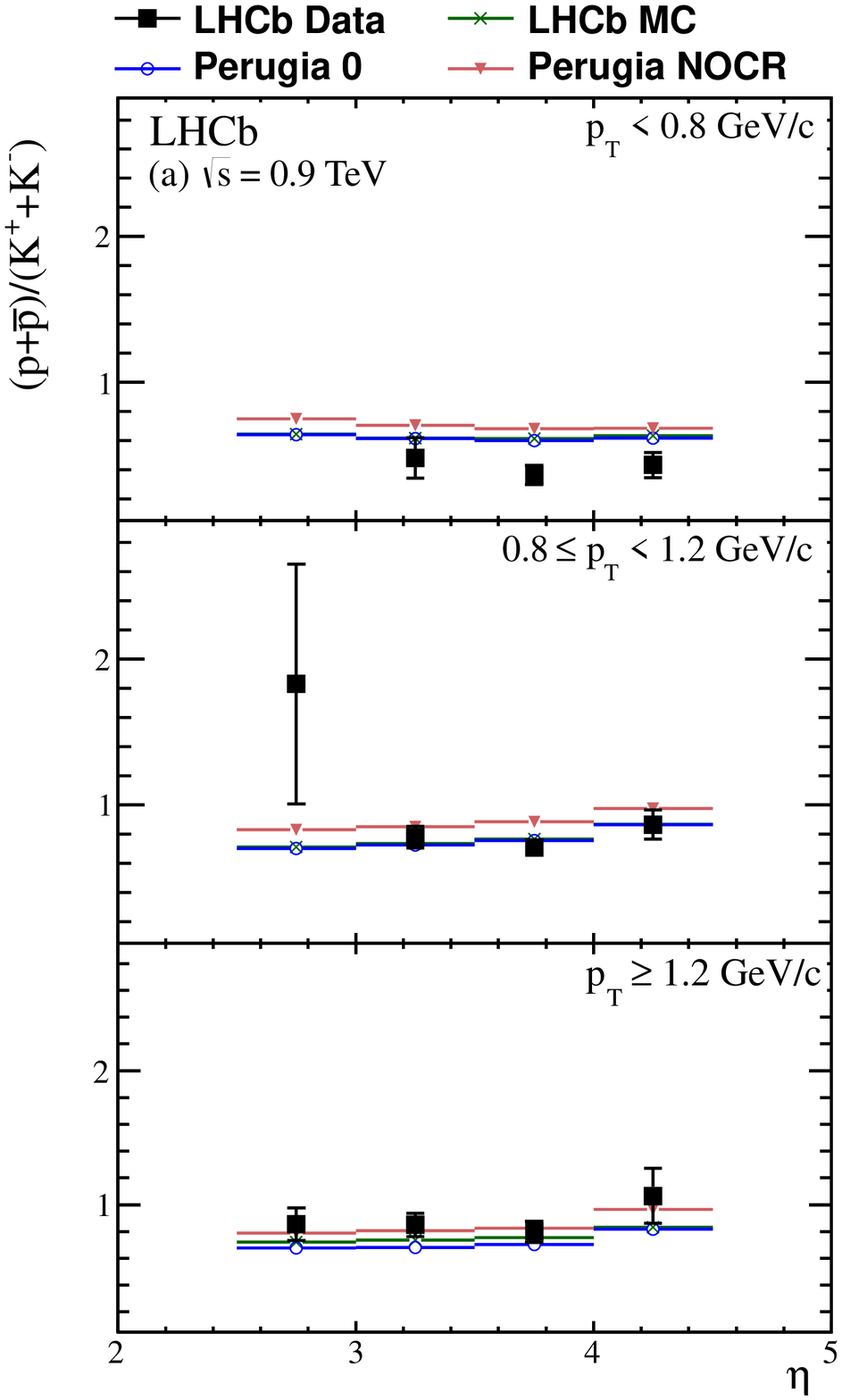}}
 \subfigure{\includegraphics[width=0.48\textwidth]{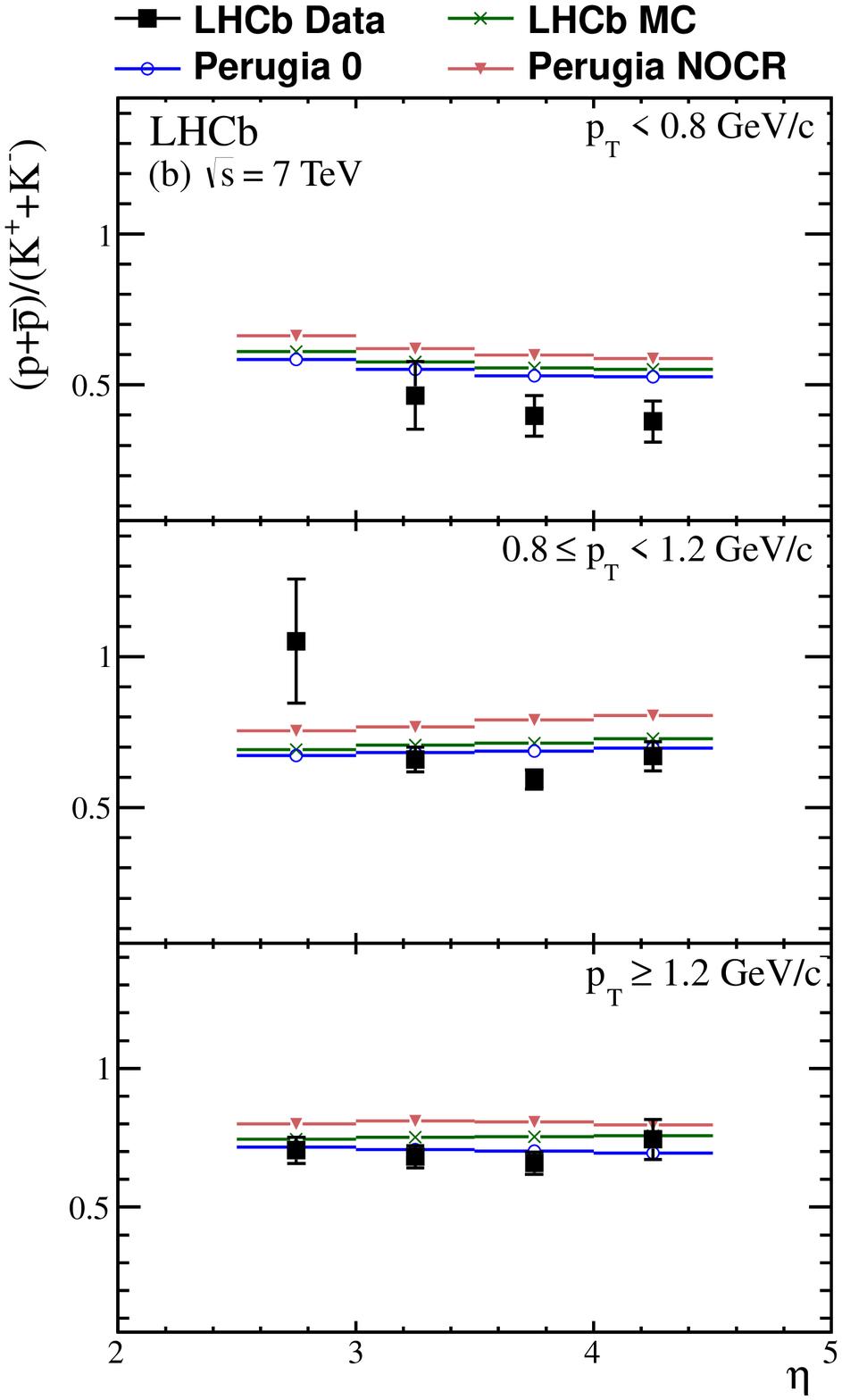}}
  \caption{\small Results for the $(p + \bar{p})/(K^+ + K^-)$ ratio at 0.9~TeV (a) and 7~TeV (b). }
 \label{fig:pK_eta_final}
 \end{center}
\end{figure}

\section{Conclusions}
\label{sec:conclusions}

Measurements have been presented of the charged-particle production ratios  $\bar{ p}/{ p}$, $K^-/K^+$, $\pi^-/\pi^+$, $(p + \bar{p})/(K^+ + K^-)$,  $(K^+ + K^-)/(\pi^+ + \pi^-)$ and $(p + \bar{p})/(\pi^+ + \pi^-)$ at both $\sqrt{s} =0.9$~TeV and
$\sqrt{s} =7$~TeV. The results at $7$~TeV are the first studies of pion, kaon and proton production to be performed at this energy. 
Comparisons have been made with several generator tunes (LHCb~MC, Perugia~0 and Perugia~NOCR).  No single tune is able to describe well all observables.  The most significant discrepancies occur for the $(p + \bar{p})/(\pi^+ + \pi^-)$ and $(K^+ + K^-)/(\pi^+ + \pi^-)$ ratios,  where the measurements are much higher than the Perugia~0 and Perugia~NOCR predictions, but lie reasonably close to the LHCb~MC expectation.   

The $\bar{p}/p$ ratio has been studied as a function of rapidity loss, $\Delta y$.  The results span the $\Delta y$ interval $3.1$ to $6.3$, and are more precise than previous measurements in this region.  Fitting a simple Regge theory inspired model to the  LHCb measurements, and those from the midrapidity region obtained by ALICE~\cite{ALICEPBARP}, yields a result with a string-junction contribution with low intercept value.

These results, together with those for related observables obtained by LHCb~\cite{LHCBV0}, will help in understanding the phenomenon of baryon-number transport, and the development of hadronisation models to improve the description of Standard Model processes in the forward region at the LHC.


\section*{Acknowledgements}

\noindent We thank Yuli Shabelski for several useful discussions.
We express our gratitude to our colleagues in the CERN accelerator
departments for the excellent performance of the LHC. We thank the
technical and administrative staff at CERN and at the LHCb institutes,
and acknowledge support from the National Agencies: CAPES, CNPq,
FAPERJ and FINEP (Brazil); CERN; NSFC (China); CNRS/IN2P3 (France);
BMBF, DFG, HGF and MPG (Germany); SFI (Ireland); INFN (Italy); FOM and
NWO (The Netherlands); SCSR (Poland); ANCS (Romania); MinES of Russia and
Rosatom (Russia); MICINN, XuntaGal and GENCAT (Spain); SNSF and SER
(Switzerland); NAS Ukraine (Ukraine); STFC (United Kingdom); NSF
(USA). We also acknowledge the support received from the ERC under FP7
and the Region Auvergne.


\appendix

\vspace*{1.0cm}

{\noindent\bf\Large Appendix}

\section{Tables of results}
\label{sec:tabresults}

The results for the same-particle ratios, including the rapidity to which the events in each pseudorapidity bin correspond, are given in Tables~\ref{tab:ratio_final_result_sum_proton},~\ref{tab:ratio_final_result_sum_kaon} and~\ref{tab:ratio_final_result_sum_pion}.  The results for the different-particle ratios can be found in Tables~\ref{tab:ratio_final_result_sum_ppi}, \ref{tab:ratio_final_result_sum_Kpi} and \ref{tab:ratio_final_result_sum_pK}.

\begin{sidewaystable}
\begin{center}
\caption{\small Results for the $\bar{p}/p$ ratio with statistical and systematic uncertainties,  as a function of \pt and $\eta$.  Also shown is the mean rapidity, $y$, and RMS spread for the sample in each $\eta$ bin.} \label{tab:ratio_final_result_sum_proton} 
\begin{tabular}{l|cc|cc|cc}
 & \multicolumn{2}{c|}{$\pt < 0.8$ GeV/$c$} & \multicolumn{2}{c|}{$0.8\le \pt<1.2$ GeV/$c$} & \multicolumn{2}{c}{$\pt \ge1.2$ GeV/$c$}\\
& $y$ (RMS) & Ratio &  $y$ (RMS) & Ratio &  $y$ (RMS) & Ratio \\
\hline
$\sqrt{s} = 0.9$~TeV & & & & & & \\ \cline{1-1}
$2.5 < \eta<3.0$ & -- & --                                                                             & $2.42$ $(0.24)$ & ${1.107 \pm 0.020 \pm 0.349}$ & $2.63$ $(0.16)$ & ${0.794 \pm 0.015 \pm 0.089}$\\
$3.0 \le \eta<3.5$ & $2.58$ $(0.27)$ & ${0.751 \pm 0.011 \pm 0.163}$  & $2.96$ $(0.25)$ &${0.684 \pm 0.010 \pm 0.049}$  & $3.08$ $(0.23)$ & ${0.614 \pm 0.010 \pm 0.047}$\\
$3.5\le \eta<4.0$ & $2.96$ $(0.11)$ & ${0.729 \pm 0.007 \pm 0.040}$  & $3.40$ $(0.22)$ &${0.576 \pm 0.007 \pm 0.032}$  & $3.56$ $(0.24)$ & ${0.456 \pm 0.009 \pm 0.033}$\\
$4.0\le \eta<4.5$ & $3.34$ $(0.24)$ & ${0.660 \pm 0.009 \pm 0.046}$  & $3.87$ $(0.14)$ &${0.451 \pm 0.009 \pm 0.038}$  & $4.02$ $(0.25)$  & ${0.328 \pm 0.010 \pm 0.049}$\\
\hline
$\sqrt{s} = 7$~TeV & & & & & &  \\  \cline{1-1}
$2.5 < \eta<3.0$ & -- & --                                                                              & $2.41$ $(0.25)$ &${1.181 \pm 0.020 \pm 0.195}$  &  $2.63$ $(0.16)$ &${0.880 \pm 0.009 \pm 0.039}$\\
$3.0\le\eta<3.5$ &  $2.55$ $(0.27)$  & ${0.734 \pm 0.011 \pm 0.124}$ & $2.98$ $(0.25)$ &${0.942 \pm 0.011 \pm 0.036}$  &  $3.12$ $(0.22)$ &${0.905 \pm 0.008 \pm 0.026}$\\
$3.5\le\eta<4.0$ &  $2.96$ $(0.09)$  & ${1.015 \pm 0.009 \pm 0.037}$ & $3.40$ $(0.23)$ &${0.916 \pm 0.007 \pm 0.022}$  &  $3.59$ $(0.24)$ &${0.903 \pm 0.008 \pm 0.023}$\\
$4.0\le\eta<4.5$ &  $3.34$ $(0.21)$  & ${0.957 \pm 0.010 \pm 0.051}$ & $3.86$ $(0.19)$ &${0.906 \pm 0.010 \pm 0.039}$  &  $4.06$ $(0.25)$  &${0.831 \pm 0.010 \pm 0.050}$\\
\end{tabular}
\end{center}
\end{sidewaystable}

\begin{sidewaystable}
\begin{center}
\caption{\small Results for the $K^-/K^+$ ratio with statistical and systematic uncertainties, as a function of \pt and $\eta$.  Also shown is the mean rapidity, $y$, and RMS spread for the sample in each $\eta$ bin.} \label{tab:ratio_final_result_sum_kaon} 
\begin{tabular}{l|cc|cc|cc}
 & \multicolumn{2}{c|}{$\pt < 0.8$ GeV/$c$} & \multicolumn{2}{c|}{$0.8\le \pt<1.2$ GeV/$c$} & \multicolumn{2}{c}{$\pt \ge1.2$ GeV/$c$}\\
& $y$ (RMS) & Ratio &  $y$ (RMS) & Ratio &  $y$ (RMS) & Ratio \\
\hline
$\sqrt{s} = 0.9$~TeV & & & & & & \\ \cline{1-1}
$2.5 < \eta<3.0$  & --     & --                                                                          & $2.65$ $(0.19)$ & ${0.870 \pm 0.010 \pm 0.267}$  &  $2.69$ $(0.14)$  & ${0.936 \pm 0.013 \pm 0.069}$\\
$3.0\le \eta<3.5$  & $2.99$ $(0.25)$  & ${0.834 \pm 0.007 \pm 0.069}$  & $3.12$ $(0.21)$  & ${0.847 \pm 0.009 \pm 0.040}$ &  $3.18$ $(0.15)$  & ${0.783 \pm 0.011 \pm 0.037}$\\
$3.5\le \eta<4.0$  & $3.32$ $(0.25)$  & ${1.001 \pm 0.007 \pm 0.064}$  & $3.62$ $(0.22)$  & ${0.792 \pm 0.009 \pm 0.028}$ &  $3.70$ $(0.17)$  & ${0.723 \pm 0.012 \pm 0.031}$\\
$4.0\le \eta<4.5$  & $3.67$ $(0.18)$  & ${1.002 \pm 0.007 \pm 0.093}$  & $4.11$ $(0.25)$  & ${0.680 \pm 0.010 \pm 0.041}$ &  $4.20$ $(0.21)$  & ${0.506 \pm 0.014 \pm 0.050}$\\ 
\hline
$\sqrt{s} = 7$~TeV & & & & & &  \\  \cline{1-1}

$2.5<\eta<3.0$ & -- & --                                                                              & $2.65$ $(0.19)$   &${0.995 \pm 0.008 \pm 0.101}$ & $2.70$ $(0.13)$ & ${0.991 \pm 0.007 \pm 0.021}$\\
$3.0 \le \eta<3.5$ &  $3.02$ $(0.25)$  & ${0.992 \pm 0.006 \pm 0.063}$ & $3.12$ $(0.21)$   &${0.966 \pm 0.006 \pm 0.019}$ & $3.20$ $(0.14)$ &${0.999 \pm 0.006 \pm 0.016}$\\
$3.5\le \eta<4.0$ &  $3.34$ $(0.25)$  & ${1.062 \pm 0.005 \pm 0.040}$ & $3.62$ $(0.21)$   &${0.948 \pm 0.006 \pm 0.014}$ & $3.70$ $(0.15)$ &${0.930 \pm 0.006 \pm 0.017}$\\
$4.0 \le \eta<4.5$ &  $3.72$ $(0.22)$   & ${1.161 \pm 0.005 \pm 0.055}$ & $4.11$ $(0.23)$   &${0.898 \pm 0.006 \pm 0.025}$ & $4.21$ $(0.18)$ &${0.958 \pm 0.009 \pm 0.049}$\\
\end{tabular}
\end{center}
\end{sidewaystable} 

\begin{sidewaystable}
\begin{center}
\caption{\small Results for the $\pi^-/\pi^+$ ratio with statistical and systematic uncertainties, as a function of \pt and $\eta$.  Also shown is the mean rapidity, $y$, and RMS spread for the sample in each $\eta$ bin.} \label{tab:ratio_final_result_sum_pion} 
\begin{tabular}{l|cc|cc|cc}
 & \multicolumn{2}{c|}{$\pt < 0.8$ GeV/$c$} & \multicolumn{2}{c|}{$0.8\le \pt<1.2$ GeV/$c$} & \multicolumn{2}{c}{$\pt \ge1.2$ GeV/$c$}\\
& $y$ (RMS) & Ratio &  $y$ (RMS) & Ratio &  $y$ (RMS) & Ratio \\
\hline
$\sqrt{s} = 0.9$~TeV & & & & & & \\ \cline{1-1}
$2.5 < \eta<3.0$ & --  & --                                                                          &  $2.74$ $(0.07)$ &${0.987 \pm 0.010 \pm 0.013}$ &  $2.75$ $(0.05)$ & ${0.970 \pm 0.016 \pm 0.014}$\\
$3.0\le\eta<3.5$ & $3.23$ $(0.09)$ &${0.979 \pm 0.005 \pm 0.010}$ &  $3.23$ $(0.07)$& ${0.971 \pm 0.011 \pm 0.010}$ &  $3.24$ $(0.05)$ & ${0.926 \pm 0.017 \pm 0.014}$\\
$3.5\le\eta<4.0$ & $3.71$ $(0.15)$ &${0.968 \pm 0.004 \pm 0.011}$ &  $3.75$ $(0.08)$& ${0.951 \pm 0.012 \pm 0.010}$ &  $3.75$ $(0.05)$ & ${0.871 \pm 0.019 \pm 0.012}$\\
$4.0\le\eta<4.5$ & $4.15$ $(0.24)$ &${0.929 \pm 0.004 \pm 0.017}$ &  $4.30$ $(0.10)$& ${0.971 \pm 0.016 \pm 0.019}$ &  $4.30$ $(0.07)$ & ${0.816 \pm 0.025 \pm 0.029}$\\ \hline
$\sqrt{s} = 7$~TeV & & & & & &  \\  \cline{1-1}
$2.5 < \eta<3.0$ & -- & --                                                                             &  $2.74$ $(0.07)$   & ${1.002 \pm 0.007 \pm 0.006}$ &  $2.74$ $(0.04)$    & ${1.015 \pm 0.010 \pm 0.005}$\\
$3.0\le\eta<3.5$ & $3.23$ $(0.09)$  & ${1.011 \pm 0.004 \pm 0.006}$ &  $3.24$ $(0.07)$   & ${0.998 \pm 0.007 \pm 0.004}$ &  $3.24$ $(0.04)$    & ${0.998 \pm 0.010 \pm 0.004}$\\
$3.5\le\eta<4.0$ & $3.70$ $(0.14)$  & ${1.002 \pm 0.003 \pm 0.006}$ &  $3.74$ $(0.07)$   & ${1.003 \pm 0.008 \pm 0.004}$ &  $3.75$ $(0.05)$    & ${1.000 \pm 0.011 \pm 0.005}$\\
$4.0\le\eta<4.5$ & $4.14$ $(0.22)$  & ${0.976 \pm 0.003 \pm 0.006}$ &  $4.26$ $(0.08)$   & ${0.998 \pm 0.009 \pm 0.008}$ &  $4.26$ $(0.05)$    & ${0.974 \pm 0.012 \pm 0.017}$\\
\end{tabular}
\end{center}
\end{sidewaystable}

\begin{table}
\begin{center}
\caption{\small Results for the $(p + \bar{p})/(\pi^+ + \pi^-)$  ratio with statistical and systematic uncertainties, as a function of \pt and $\eta$.} 
\label{tab:ratio_final_result_sum_ppi}
\begin{tabular}{l|c|c|c}
 & $\pt <0.8$  GeV/$c$ & $0.8\le \pt<1.2$  GeV/$c$ & $\pt \ge1.2$  GeV/$c$ \\
\hline
$\sqrt{s} = 0.9$~TeV & & &  \\ \cline{1-1}
$2.5 < \eta<3.0$ & --  & ${0.328 \pm 0.007 \pm 0.104}$ & ${0.300 \pm 0.008 \pm 0.034}$\\
$3.0\le\eta<3.5$ & ${0.086 \pm 0.001 \pm 0.021}$ & ${0.208 \pm 0.004 \pm 0.016}$ & ${0.272 \pm 0.007 \pm 0.023}$\\
$3.5\le\eta<4.0$ & ${0.062 \pm 0.001 \pm 0.008}$ & ${0.175 \pm 0.003 \pm 0.011}$ & ${0.252 \pm 0.007 \pm 0.020}$\\
$4.0\le\eta<4.5$ & ${0.076 \pm 0.001 \pm 0.010}$ & ${0.233 \pm 0.006 \pm 0.022}$ & ${0.301 \pm 0.013 \pm 0.047}$\\
\hline
$\sqrt{s} = 7$~TeV & & &  \\ \cline{1-1}
$2.5 < \eta<3.0$ & -- & ${0.235 \pm 0.004 \pm 0.039}$ & ${0.262 \pm 0.004 \pm 0.014}$\\
$3.0\le\eta<3.5$ & ${0.085 \pm 0.001 \pm 0.017}$ & ${0.174 \pm 0.002 \pm 0.009}$ & ${0.245 \pm 0.003 \pm 0.011}$\\
$3.5\le\eta<4.0$ & ${0.069 \pm 0.001 \pm 0.008}$ & ${0.156 \pm 0.002 \pm 0.006}$ & ${0.242 \pm 0.003 \pm 0.010}$\\
$4.0\le\eta<4.5$ & ${0.051 \pm 0.001 \pm 0.007}$ & ${0.184 \pm 0.003 \pm 0.010}$ & ${0.244 \pm 0.004 \pm 0.017}$\\
\end{tabular}
\end{center}
\end{table}

\begin{table}
\begin{center}
\caption{\small Results for the $(K^+ + K^-)/(\pi^+ + \pi^-)$  ratio with statistical and systematic uncertainties,  as a function of \pt and $\eta$.} 
\label{tab:ratio_final_result_sum_Kpi}
\begin{tabular}{l|c|c|c} 
 & $\pt <0.8$ GeV/$c$ & $0.8\le \pt<1.2$ GeV/$c$ & $\pt \ge 1.2$ GeV/$c$\\
\hline
$\sqrt{s} = 0.9$~TeV & & &  \\ \cline{1-1}
$2.5 < \eta<3.0$ & --  & ${0.184 \pm 0.003 \pm 0.056}$ & ${0.351 \pm 0.008 \pm 0.028}$\\
$3.0\le\eta<3.5$ & ${0.180 \pm 0.002 \pm 0.026}$ & ${0.267 \pm 0.004 \pm 0.015}$ & ${0.319 \pm 0.008 \pm 0.018}$\\
$3.5\le\eta<4.0$ & ${0.171 \pm 0.001 \pm 0.023}$ & ${0.247 \pm 0.004 \pm 0.011}$ & ${0.314 \pm 0.009 \pm 0.017}$\\
$4.0\le\eta<4.5$ & ${0.173 \pm 0.001 \pm 0.025}$ & ${0.268 \pm 0.006 \pm 0.018}$ & ${0.281 \pm 0.012 \pm 0.031}$\\
\hline
$\sqrt{s} = 7$~TeV & & &  \\ \cline{1-1}
$2.5 < \eta<3.0$ & -- & ${0.224 \pm 0.002 \pm 0.024}$ & ${0.371 \pm 0.004 \pm 0.014}$\\
$3.0\le\eta<3.5$ & ${0.181 \pm 0.001 \pm 0.024}$ & ${0.263 \pm 0.003 \pm 0.010}$ & ${0.357 \pm 0.004 \pm 0.012}$\\
$3.5\le\eta<4.0$ & ${0.173 \pm 0.001 \pm 0.021}$ & ${0.262 \pm 0.003 \pm 0.009}$ & ${0.367 \pm 0.005 \pm 0.013}$\\
$4.0\le\eta<4.5$ & ${0.131 \pm 0.001 \pm 0.016}$ & ${0.275 \pm 0.003 \pm 0.011}$ & ${0.328 \pm 0.005 \pm 0.020}$\\
\end{tabular}
\end{center}
\end{table}

\begin{table}
\begin{center}
\caption{\small Results for the $(p + \bar{p})/(K^+ + K^-)$ ratio with statistical and systematic uncertainties,  as a function of \pt and $\eta$.} 
\label{tab:ratio_final_result_sum_pK}
\begin{tabular}{l|c|c|c}
 & $\pt <0.8$ GeV/$c$ & $0.8\le \pt<1.2$ GeV/$c$& $\pt \ge1.2$ GeV/$c$\\
\hline
$\sqrt{s} = 0.9$~TeV & & &  \\ \cline{1-1}
$2.5 < \eta<3.0$ & --  & ${1.831 \pm 0.039 \pm 0.822}$ & ${0.855 \pm 0.020 \pm 0.119}$\\
$3.0\le\eta<3.5$ & ${0.481 \pm 0.008 \pm 0.139}$ & ${0.779 \pm 0.014 \pm 0.073}$ & ${0.851 \pm 0.019 \pm 0.084}$\\
$3.5\le\eta<4.0$ & ${0.363 \pm 0.004 \pm 0.066}$ & ${0.709 \pm 0.012 \pm 0.055}$ & ${0.799 \pm 0.021 \pm 0.076}$\\
$4.0\le\eta<4.5$ & ${0.433 \pm 0.007 \pm 0.086}$ & ${0.865 \pm 0.021 \pm 0.097}$ & ${1.067 \pm 0.045 \pm 0.200}$\\
\hline
$\sqrt{s} = 7$~TeV & & &  \\ \cline{1-1}
$2.5 < \eta<3.0$ & --  & ${1.051 \pm 0.020 \pm 0.204}$ & ${0.705 \pm 0.009 \pm 0.046}$\\
$3.0\le\eta<3.5$ & ${0.465 \pm 0.008 \pm 0.111}$ & ${0.660 \pm 0.009 \pm 0.039}$ & ${0.682 \pm 0.007 \pm 0.038}$\\
$3.5\le\eta<4.0$ & ${0.398 \pm 0.004 \pm 0.067}$ & ${0.593 \pm 0.006 \pm 0.031}$ & ${0.659 \pm 0.007 \pm 0.037}$\\
$4.0\le\eta<4.5$ & ${0.379 \pm 0.004 \pm 0.068}$ & ${0.671 \pm 0.009 \pm 0.046}$ & ${0.744 \pm 0.011 \pm 0.069}$\\
\end{tabular}
\end{center}
\end{table}

\clearpage


\bibliographystyle{LHCb}
\bibliography{partRatio}

\end{document}